\documentclass[12pt,draftcls,onecolumn]{IEEEtran}
\usepackage[utf8]{inputenc}
\usepackage{amsmath,mathtools}
\usepackage{amsthm}
\usepackage{amsfonts}
\usepackage{amssymb}
\usepackage{easyReview} 
\usepackage{soulutf8} 
\DeclarePairedDelimiter\abs{\lvert}{\rvert}
\DeclarePairedDelimiter\norm{\lVert}{\rVert}
\usepackage{graphicx}
\usepackage{float}
\usepackage{tikz}
\usepackage{changebar}
\usepackage{authblk}
\usetikzlibrary{shapes,snakes,patterns}
\usepackage{epstopdf}
\renewcommand{\vec}[1]{\mathbf{#1}}
\usepackage[justification=centering]{caption}
\usepackage{enumerate} 
\usepackage{cite}
\newtheorem{theorem}{Theorem}[]

\usepackage{suffix}
\usepackage{enumitem}
\usepackage{xfrac}
\usepackage{relsize}
\usepackage{dsfont}
\usepackage{bm}
\usepackage{array}
\newcolumntype{C}[1]{>{\centering\let\newline\\\arraybackslash\hspace{0pt}}m{#1}}
\usepackage{algorithm,algorithmic}
\theoremstyle{definition}

\usepackage[list=true]{subcaption}
\begin{document}
	\bstctlcite{IEEEexample:BSTcontrol}
	\title{Joint Power-control and Antenna Selection in User-Centric Cell-Free Systems with Mixed Resolution ADC}
	\author{Shashank Shekhar$^{1}$\thanks{1. Shashank Shekhar, Athira Subhash and Sheetal Kalyani are with the Dept. of Electrical Engineering, Indian Institute of Technology, Madras, India.
			(Emails: \{ee17d022@smail, ee16d027@smail, skalyani@ee\}.iitm.ac.in).}, Athira Subhash$^{1}$, Muralikrishnan Srinivasan$^{2}$\thanks{2. Muralikrishnan Srinivasan is with the Dept. of Electrical Engineering, Chalmers University of Technology, Gothenburg, Sweden.(Email: mursri@chalmers.se) }, and Sheetal Kalyani$^{1}$  
	}
	
	\maketitle
	\begin{abstract}
		In this paper, we propose a scheme for the joint optimization of the user transmit power and the antenna selection at the access points (AP)s of a user-centric cell-free massive multiple-input-multiple-output (UC CF-mMIMO) system. We derive an approximate expression for the achievable uplink rate of the users in a UC CF-mMIMO system in the presence of a mixed analog-to-digital converter (ADC) resolution profile at the APs. Using the derived approximation, we propose to maximize the uplink sum-rate of UC CF-mMIMO systems subject to energy constraints at the APs. An alternating-optimization solution is proposed using binary particle swarm optimization (BPSO) and successive convex approximation (SCA). We also study the impact of various system parameters on the performance of the system.
	\end{abstract}
	

	\section{Introduction}
	The concept of massive multiple-input-multiple-output (mMIMO) systems with a vast number of transmitting or receiving antennas or both have received much theoretical and practical attention over the last few years \cite{marzetta2016fundamentals}. The idea of a large number of co-located antennas has been then extended to distributed mMIMO systems \cite{zhou2003distributed}, and the performance improvements achievable from such distributed transceivers have been studied under different system models in papers like \cite{madhow2014distributed,venkatesan2007network, Ngo2017:CellFree,scalablecellfree}. In a conventional cellular communication system, geographical areas are divided into non-overlapping regions called cells, and all users within a cell are served by a single base station (BS). Massive densification of such cells, popularly termed as small cell communication, has been proven to be one viable solution for achieving high energy efficiency (EE) \cite{bjornson2016deploying}. However, the future demands of wireless communication systems are multi-faceted and more challenging to achieve. Hence, the small cell communication systems may not cater to all of them.  
	\par Recently, the idea of cell-free (CF) communication has attracted much attention \cite{Ngo2017:CellFree}. A CF-mMIMO system comprises a large number of access points (AP)s simultaneously serving a smaller (compared to the number of APs) number of users with the help of a central processing unit (CPU) supervising the whole communication system. CF communication has been demonstrated to improve the performance over the small cell schemes and thus is a promising technology capable of achieving the demands of 5G and 6G communication standards.
	\begin{table}[]
		\centering
		\begin{tabular}{|C{1.5cm}|C{1.5cm}|C{1cm}|C{4cm}|C{1.5cm}|C{4cm}|}
			\hline
			Reference & User-centric  & ADC / Mixed ADC & Objective & Energy constraint & Optimization methodology\\
			\hline
			\cite{buzzi2019user} & $\checkmark$  & \texttimes & sum-rate \& minimum rate & \texttimes & Successive Lower-Bound Maximization \\
			\hline
			\cite{buzzi2017downlink} & $\checkmark$  & \texttimes & sum-rate \& minimum rate & \texttimes & Successive Lower-Bound Maximization \\
			\hline
			\cite{alonzo2019energy} & $\checkmark$  & \texttimes & Ratio of sum-rate to energy consumption & \texttimes & Successive Lower-Bound Maximization \\
			\hline
			\cite{li2019power}& $\checkmark$  & \texttimes & minimum rate & \texttimes & Bisection method\\
			\hline
			\cite{hu2018rate}& \texttimes & $\checkmark$ & sum-rate & \texttimes & lower bound maximization\\
			\hline
			\cite{hu2019cell}& \texttimes  & $\checkmark$ & sum-rate \& minimum rate & \texttimes & lower bound maximization \& Bisection method\\
			\hline
			\cite{MixedADCCellFree}& \texttimes  & $\checkmark$ & minimum rate & \texttimes & geometric programming\\
			\hline
			This paper & $\checkmark$  & $\checkmark$ & sum-rate & $\checkmark$ & alternating optimization using SCA-GP and BPSO \\
			\hline
		\end{tabular}
		\caption{Comparison of this work with existing literature}
		\label{tab:LiteratureTable}
	\end{table}

	\par Several papers in the literature like \cite{zhang2019cell,beyond_5g,athreya2020beyond,nayebi2017precoding,yang2018energy,gopi2021cooperative,ozdogan2019performance,mai2018cell,shekhar2021outage} have investigated multiple aspects of CF systems. Detailed surveys and studies on CF systems are available in papers such as \cite{zhang2019cell}, \cite{beyond_5g} and references therein. The authors of \cite{nayebi2017precoding} show that CF-mMIMO systems can achieve higher rates  (in terms of 5$\%$-outage and minimum rate) when compared to small-cell users even when power allocation is applied more frequently in the small cell scheme. Furthermore, the authors of  \cite{yang2018energy} compare the EE and spectral efficiency (SE) of CF systems with those of single-cell mMIMO systems. They also prove that CF systems can perform better with good power allocation strategies like the max-min power control schemes, though time-consuming. They rightly emphasize that the utility of the resource allocation algorithms depends both on the quality of performance and the implementation complexity. The uplink and downlink SE of a CF-mMIMO system under Rician channel fading conditions have been studied in \cite{ozdogan2019performance}. The authors also evaluate the performance losses incurred in the uplink and downlink channels due to the unavailability of the phase of the line of sight (LoS) paths. The performance of CF systems with multi-antenna APs and multi-antenna user equipment (UE) is studied in \cite{mai2018cell}. Furthermore, a closed-form expression for the achievable downlink  SE  considering the availability of imperfect channel state information (CSI), non-orthogonal pilots, and power control is provided. Recently, approximate outage probability (OP) expressions are derived for uplink CF-mMIMO systems using the dimension reduction method for multifold integration in \cite{shekhar2021outage}. 
	\par In a conventional CF system, all the users are served by all the APs. However, in large networks, each user is physically close to only a finite set of APs. Hence, the authors of \cite{buzzi2017cell} have introduced the concept of a user-centric (UC) virtual cell approach to CF-mMIMO, wherein each user is served only by a limited number of AP's \cite{demir2021foundations}. UC CF systems can be implemented with lesser backhaul overhead and were demonstrated to outperform the conventional CF systems in terms of achievable rate-per-user for the vast majority of the users in the network \cite{buzzi2017cell}. UC CF-mMIMO architecture with multiple antennas at both the APs and users has been studied in \cite{buzzi2019user}. The authors of \cite{alonzo2017cell} have investigated the performance of UC CF-mMIMO at millimeter-wave frequencies. Downlink power control strategies to maximize the sum-rate or the minimum rate for a UC CF-mMIMO system are explored in \cite{buzzi2017downlink}. Another essential aspect of UC CF-mMIMO systems is the user assignment strategies. One such strategy which ensures that at least one AP serves each user is discussed in \cite{li2019power}.
	We provide a summary and contrast our contributions to the existing literature in Table \ref{tab:LiteratureTable}.
	\par The usage of a large number of antennas in both mMIMO and CF-mMIMO systems can significantly increase the power consumption from radio frequency (RF) circuits, and digital signal processing (DSP) units \cite{bai2015energy, rusek2012scaling, yuan2017distributed}. Moreover, the power consumption of an analog-to-digital converter (ADC) is known to scale linearly with the signal bandwidth roughly and exponentially with the quantization bit \cite{zhang2017performance}. For example, the power dissipation of an eight-bit ADC with a sampling rate of $20$ Giga-samples per second is $10~Watt$ \cite{murmann2008d}. Hence, the more the number of antennas at the AP, the higher the power consumption at the AP. Therefore, energy efficiency is a significant performance metric in wireless communications, especially in CF-mMIMO systems that employ many antennas.
	
	\par The ADC power consumption could be reduced by reducing the bit resolution or up-scaling the number of antennas, or keeping the number of users constant \cite{sarajlic2017low}. However, on the other hand, the reduction in the spectral efficiency due to low-bit resolution ADCs is also a severe concern \cite{liu2019energy}. One solution to mitigate the loss would be to have mixed resolution ADC architecture at the APs.
	\par The performance of mixed resolution ADC architecture has been extensively studied in the case of cellular mMIMO systems. For example, the authors of \cite{zhang2016spectral,srinivasan2019analysis} have derived an approximate expression of the uplink SE and outage probability  of mMIMO systems with MRC receivers using the additive quantization noise model (AQNM) for modeling the imperfections in the received signal due to low-bit resolution ADCs, respectively. Similarly, an approximate analytical expression for the uplink achievable rate of an mMIMO system with finite-precision ADCs and maximal ratio combining (MRC) receivers has been studied in \cite{fan2015uplink}. The authors of \cite{srinivasan2019joint} have considered the joint power control and resource allocation for a device-to-device (D2D) underlay cellular system with a multi-antenna BS employing ADCs with different resolutions. Similarly, the uplink achievable SE of multi-user distributed mMIMO with mixed-ADC receivers has been studied in \cite{yuan2017distributed}. Their study has concluded that the distributed mixed-ADC architecture is an energy-efficient design capable of outperforming the centralized mMIMO system. Furthermore, the architecture can achieve large throughputs by deploying a large number of low-resolution remote radio heads (L-RRHs).

	\par Inspired by these results, the performance of CF systems with low-bit resolution ADCs has also been analyzed in many works. The authors of \cite{zhang2019performance} have considered a CF-system with multi-antenna users and multi-antenna BS employed with low-bit resolution ADCs. Their observations reveal that with an appropriate choice of quantization bits, the CF-mMIMO with low-bit resolution ADCs can achieve a better SE-EE trade-off when compared to the perfect high-bit resolution ADC counterpart. Furthermore, the authors of \cite{hu2019cell} have proved that the asymptotic achievable rate (when the number of APs is large) of users in the CF system converges to a finite limit independent of the ADC resolutions. They prove that the ADC resolution of the user alone determines this limit. An optimal ADC resolution bit allocation scheme that maximizes the sum-rate subject to the total ADC resolution bit constraint has been developed in \cite{hu2018rate}. 

	\par All the prior art studying/optimizing the performance of low-bit resolution ADCs at the APs assumes the basic CF-system model where all the APs serve all the users. None of the works, even for a basic CF-system, let alone a UC CF-system develop antenna selection schemes to balance EE with SE. An antenna selection scheme is crucial because some antennas attached to ADCs with certain-bit resolutions can contribute very little to improving spectral efficiency. At the same time, the power required to activate the RF chain of that antenna becomes a burden penalizing the system's total energy efficiency (EE).  Therefore, posing the trade-off as an optimization problem and solving it would determine the most feasible solution.

	\par Hence, our paper considers a UC CF-mMIMO system with mixed resolution ADC architecture at the APs and single-antenna users. Here, few antennas at the APs are assumed to be connected to RF chains with high-bit resolution ADCs, and the rest are considered to have access to low-bit resolution ADCs only. In a usual antenna selection problem, all the antennas are assumed to be identical. In contrast, the antennas are not similar here because of the different ADC resolutions. In this paper, we propose a scheme for the joint optimization of the user transmit power and the antenna selection at the APs to maximize the uplink sum rate utility while satisfying a maximum energy consumption constraint at the APs. The main contributions of this paper are summarized as follows:
	\begin{itemize}
		\item We derive a closed-form lower bound for the achievable uplink rate of the users in a UC CF-mMIMO system in the presence of a mixed ADC resolution profile at the APs, using the popular use-and-forget bound.
		\item For sum-rate maximization, we propose an optimization technique that alternates between two sub-problems corresponding to the optimization of power coefficients and antenna-selection coefficients, respectively. For power control, we use the successive convex approximation (SCA) to relax the non-convex objective into a convex objective and use geometric programming (GP) to solve the relaxed convex sub-problem. Binary particle swarm optimization (BPSO) is used for optimizing the antenna-selection coefficients. 
		\item We also study the impact of various system parameters such as number of APs, number of users, number of users served by an AP, ratio of high-resolution to total number of antenna and permissible energy consumption on the performance of the system.
	\end{itemize}
	\subsubsection*{Organization} The rest of the paper is organized as follows. Section \ref{sec:UCCF_SysModel} describes the system model under consideration. Next, in Section \ref{sec:UCCF_Rate}, we evaluate the lower bound on the achievable uplink rate. In Section \ref{sec:UCCF_OptProb}, we formulate the optimization problem to maximize the achievable uplink sum-rate under constraints on energy consumption of the APs and the power coefficient of the users. We also discuss the proposed solution in this section. Section \ref{sec:UCCF_Results} presents the simulation results and finally, concluding remarks are presented in Section \ref{sec:UCCF_Conclusion}.
	\subsubsection*{Notation} Following notations are used in this paper: $\mathbf{X}$ denotes a matrix, $\mathbf{x}$ denotes a vector and $x$ denotes a scalar. Also, $\left(\cdot\right)^{H}$ and $\operatorname{Tr}\left(\cdot\right)$ represents the conjugate transpose and trace operator, respectively. $\Vert \cdot \Vert$ denotes the euclidean norm operator. $\operatorname{diag}\left( \mathbf{X}\right)$ returns a diagonal matrix with diagonal elements of $\mathbf{X}$ and, $\operatorname{diag}\left( \mathbf{x}\right)$ returns a diagonal matrix with the elements of vector $\mathbf{x}$. $\mathbf{I}_{N}$ and $ \mathbf{0}_{N} $ denotes the $N \times N$ identity and zero matrix, respectively. $\mathcal{U}[a,b]$ represents the uniform distribution over the support $[a,b]$.   
	\section{System Model}\label{sec:UCCF_SysModel}
	A CF-mMIMO system with $ M $ APs and $ K $ users is considered, where $ M \gg K $. Each AP is equipped with $ N $ antennas and the users are equipped with a single antenna each. We have considered a mixed ADC structure, where the ADCs connected to the antennas have different resolution \cite{MixedADCCellFree},\cite{MixedADCCellFreeRician}. Out of the $ N $ antennas at each of the AP, $ N_{1} $ antennas are connected with low-bit resolution ADCs and the remaining $ N_{2} = N- N_1$ antennas are connected with high-bit resolution ADCs. Let $ b_{m}^{i} $ denotes the resolution of ADC connected to the $ i $th antenna of the $ m $th AP. Unlike \cite{MixedADCCellFree},\cite{MixedADCCellFreeRician} where $ b_{m}^{i}$ was considered to be same for all the $ N_{1} $ antenna, we have considered that the first $ N_{1} $ antenna of all the APs to have a resolution profile ranging form $1$-bit to $N_{1} $-bit, \textit{i.e.}, $ b_{m}^{i} = i \  \forall \ i \in \lbrace 1,2,\dots,N_{1} \rbrace \ \text{and} \  \forall \ m \in \lbrace 1,2,\dots,M \rbrace$. This particular choice of resolution profile provides more flexibility compared to the fixed low resolution profile. The channel between the $ m $th AP and the $ k $th user is modeled as a Rayleigh fading channel.  Let  $ \mathbf{g}_{mk} \in \mathbb{C}^N$, represent the channel vector  between the $ m $th AP and the $ k $th user and we have,
	\begin{align}
		\mathbf{g}_{mk} \sim \mathcal{CN}\left(\vec 0, \beta_{mk}\mathbf{I}_{N}\right),
	\end{align}
	where $ \beta_{mk}$  represents the large scale fading coefficients between the $ m $th AP and the $ k $th user. We assume that the knowledge of $ \beta_{mk}$ is available at both the AP and the UE. Instead of considering a conventional CF-mMIMO system described in \cite{Ngo2017:CellFree}, we consider a UC version of the CF-mMIMO system presented in \cite{buzzi2019user,alonzo2017cell}. In the vanilla CF-mMIMO system, all the users are served by all the APs, whereas in a UC CF-mMIMO, an AP will serve only a fixed predefined number of users, say $L$. Each AP is
		connected to the CPU, which
		is responsible for the AP cooperation, baseband processing and user-associations for the APs. We consider TDD mode of operation, where the uplink and downlink transmissions occupy non-overlapping time intervals. Let $ \tau_{c} $ be the length of coherence interval (in samples). A part of $ \tau_{c} $, say $ \tau_{p} $, will be used for uplink training and remaining time \textit{i.e.} $ \tau_{c} - \tau_{p} = \tau_{u} $ will be used for the uplink data transmission. In this paper, we focus only on the uplink EE, and hence we will not consider the downlink data transmission phase.
	
	\subsection{Uplink Training}\label{SubSec:UCCF_UT}
	In this phase, all the users simultaneously transmit their pilot sequences to the APs. Let $\sqrt{\tau_{p}} \boldsymbol{\phi}_{k}  \in \mathbb{C}^{\tau_{p} \times 1}$ be the pilot sequence transmitted by the $ k $th user, $ \forall k = 1, \dots , K $, where $ \Vert \boldsymbol{\phi}_{k} \Vert^{2} = 1 $. We can utilize the high-bit resolution ADCs in a round-robin fashion at each of the APs to ensure quality channel estimation without the quantization error caused by the low-bit resolution ADCs \cite{RoundRobin}. The minimum-mean squared error (MMSE) estimate of $\vec {g}_{mk}$, denoted by $\vec{\hat{g}}_{mk}$, is also a complex Gaussian vector $\vec{\hat{g}}_{mk} \sim \mathcal{CN}\left(\vec 0, \gamma_{mk}\mathbf{I}_{N}\right)$, where the effective channel gain of the $ k $th user at the $ m $th AP is     
	\begin{equation} \label{Eq:UCCF_Effective_channel_Gain}
		\gamma_{mk} = \frac{\tau_{p} \rho_{p} \beta_{m k}^{2}}{\tau_{p} \rho_{p} \sum_{k^{\prime}=1}^{K} \beta_{m k^{\prime}}\left|\boldsymbol{\phi}_{k^{\prime}}^{H} \boldsymbol{\phi}_{k}\right|^{2}+1},
	\end{equation} where $\rho_{p}$ is the normalized transmit signal-to-noise ratio (SNR) of each pilot symbol \cite[eq. (5)]{Total_Energy}.
	\begin{algorithm}
		\caption{UE Selection}
		\begin{algorithmic}[1]\label{Algo:UCCF_UEselection}
			\renewcommand{\algorithmicrequire}{\textbf{Input:}}
			\renewcommand{\algorithmicensure}{\textbf{Output:}}
			\REQUIRE $\gamma_{mk} \ \forall \ m = 1,\dots,M, k = 1,\dots,K $ \text{and} $L$
			\ENSURE  Set $\mathcal{K}_{m} \forall \ m = 1,\dots,M$ \text{and} $\boldsymbol{\Theta}$
			\hrule
			\vspace{2mm}
			\FOR {$m = 1$ to $M$}
			\STATE \begin{enumerate}
				\item Sort $\gamma_{mk} \forall \ k = 1,\dots,K$ in descending order.
				\item Select first $L$ user from the sorted vector. 
				\item Store the index of selected users in $\mathcal{K}_{m}$
			\end{enumerate}
			\FOR {$k = 1$ to $K$}
			\STATE $\theta_{mk} = \begin{cases} 1, k \in \mathcal{K}_{m} 
				\\ 0, k \notin \mathcal{K}_{m}
			\end{cases}$
			\ENDFOR
			\ENDFOR
			\WHILE{true}
			\IF{$\sum_{m=1}^{M}\theta_{mk} \ne 0 \ \forall k = 1,\dots,K$} 
			\STATE break
			\ELSE 
			\STATE \begin{enumerate}
				\item Find AP which has strongest connection with $k$th user
				\begin{equation*}
					m^{*} = \arg\max_{m} \gamma_{mk}  
				\end{equation*}
				\item Find user which has weakest connection with $m^{*}$th AP
				\begin{equation*}
					k^{*} = \arg\min_{k \in \mathcal{K}_{m^{*}}} \gamma_{m^{*}k}  
				\end{equation*}
				\item Reset $\theta_{m^{*}k} = 1$ and $\theta_{m^{*}k^{*}} = 0$
			\end{enumerate}
			\ENDIF
			\ENDWHILE
		\end{algorithmic}
	\end{algorithm}
	
	\subsection{Uplink Data Transmission}
	\begin{table}[]
		\centering
		\begin{tabular}{|c|c|c|c|c|c|}
			\hline
			$ b_{m}^{i} $ & $ 1 $ & $ 2 $ & $ 3 $ & $ 4
			$ & $ 5 $\\
			\hline
			$ \alpha_{m}^{i} $	& $ 0.6366 $ & $ 0.8825 $ & $ 0.96546 $ & $ 0.990503 $ & $ 0.997501 $ \\
			\hline
		\end{tabular}
		\caption{Impairment factor $ \alpha_{m}^{i} $ for ADC with $ b_{m}^{i} $ quantization bit}
		\label{tab:my_label}
	\end{table}
	In the uplink data transmission phase, all the users send their intended data symbols to APs. Let $s_{k}$ be the symbol sent by the $k$th user such that $\mathbb{E}\left[ |s_{k}|^{2} \right] = 1$. In a UC CF-mMIMO system, the data symbol of the $k$th user is decoded by only those APs serving the $k$th user. The $m$th AP will select the $L$ users with the strongest effective channel gains given by (\ref{Eq:UCCF_Effective_channel_Gain}). Let $\mathcal{K}_{m}$ denote the set of users which are served by the $m$th AP and $\theta_{mk}$ is an indicator variable to denote whether the $m$th AP serves the $k$th user (\textit{denoted by} $\theta_{mk} = 1$) or not (\textit{denoted by} $\theta_{mk} = 0$). The heuristic used for UE selection is discussed in Algorithm \ref{Algo:UCCF_UEselection} and is the same as the one used by the authors of \cite{li2019power}. We can form the sets $\mathcal{M}_{k} = \{m : k \in \mathcal{K}_{m}\}$ which is the collection of APs serving the $k$th user, using the sets $\mathcal{K}_{m}, \ \forall m = 1,\dots,M $ . The signal received at the $m$th AP is given by 
	\begin{equation}
		\tilde{\vec y}_{m} = \sqrt{\rho_{u}}\sum_{k=1}^{K} \vec g_{mk} \sqrt{\eta_{k}} s_{k} + \vec w_{m},
	\end{equation}
	where $\rho_{u}$ is the normalized uplink
	SNR, $ \eta_{k} $ is the power control coefficient of the $k$th user, and $\vec w_{m}$ is the additive complex Gaussian noise with $\vec w_{m} \sim \mathcal{CN}\left(\vec 0, \mathbf{I}_{N} \right)$. To quantify the effect of low-bit resolution ADCs, we used the additive quantization noise model (AQNM) \cite{fan2015uplink}. With AQNM, the received quantized signal will be as follows,
	\begin{equation}
		\vec y_{m} = \sqrt{\rho_{u}}\sum_{k=1}^{K}\mathbf{A}_{m} \vec g_{mk}\sqrt{\eta_{k}} s_{k} + \mathbf{A}_{m}\vec w_{m} + \vec w_{m}^{q},
	\end{equation}
	where  $ \mathbf{A}_{m} $ is a $ N \times N $ diagonal matrix with the $ i $th diagonal element denoted by  $ \alpha_{m}^{i} $, representing the impairment factor of low-bit resolution ADCs. The resolution in bits $ b_{m}^{i} $ and the corresponding impairment factor $ \alpha_{m}^{i} $ for $ b_{m}^{i} \in \lbrace 1 \dots 5\rbrace $, are given in Table \ref{tab:my_label}. For $ b_{m}^{i} > 5 $, we can use the relation 
	\begin{equation}
		\alpha_{m}^{i} = 1 - \dfrac{\pi \sqrt{3}}{2}2^{-2 b_{m}^{i}}.
	\end{equation}  
	Also, for the $ N_{2} $ antennas which are connected to the high-bit resolution ADCs, $ \alpha_{m}^{i} = 1$. Note, $\vec w_{m}^{q}$ is the quantization noise at the $m$th AP which is modeled as an independent additive Gaussian noise with zero mean and the covariance matrix for a given channel realization is given by
	\begin{equation}
		\mathbb{E}\left[\vec w_{m}^{q} (\vec w_{m}^{q})^{H}\right] = \mathbf{A}_{m} \left(\mathbf{I}_{N} - \mathbf{A}_{m}\right)\operatorname{diag}(\mathbb{E}\left[\tilde{\vec y}_{m} \tilde{\vec y}_{m}^{H}\big| \vec g_{mk}\right]).
	\end{equation}
	In other words, 
	\begin{equation}
		\vec w_{m}^{q} \sim \mathcal{CN}\left(\vec 0,  \mathbf{A}_{m} \left(\mathbf{I}_{N} - \mathbf{A}_{m}\right)\operatorname{diag}\left(\rho_{u}\vec G_{m}\vec P \vec G_{m}^{H} + \mathbf{I}_{N}\right)\right),
	\end{equation}
	where, $\vec G_{m} = \left[\vec g_{m1} \dots \vec g_{mK} \right]$ and $\vec P = \operatorname{diag}\left(\boldsymbol{\eta} \right)$ with $\boldsymbol{\eta} = \left[\eta_{1},\dots,\eta_{K} \right]$. As we know that each user will be served only by a subset of APs, the received signal $ \vec y_{m}$ will be processed using MRC \textit{i.e.,}  multiplied with  $\hat{\vec g}_{mk}^{H}$ and transmitted to the CPU by only those APs via the backhaul network. Let $ \vec D_{mk}= \operatorname{diag}(d_{mk}^1,..., d_{mk}^N)$ be the $ N \times N $ diagonal matrix which represent the association of the $ n $th antenna of the $m$th AP and the $ k $th user. The diagonal entries $d_{mk}^n \in \{0,1\}$, for  $n=1,.., N$, $ m = 1,\dots, M $ and $ k = 1,\dots, K $ takes the value $ 1 $ if the $ n $th antenna of $ m $th AP can decode the signal from $ k $th user and $ d_{mk}^n = 0$ otherwise. If $ m \notin \mathcal{M}_{k}$, then $\vec D_{mk} = \mathbf{0}_{N} $. Hence, the received signal at CPU after MRC will be,
	\begin{equation}\label{Eq:UCCF_sk}
		\begin{aligned}
			\hat{s_k} &= \sum_{m \in \mathcal{M}_{k}} \hat{\vec g}_{mk}^H \vec D_{mk} \vec y_{m}, \\
			&= \sum_{m = 1}^{M} \hat{\vec g}_{mk}^H \vec D_{mk} \vec y_{m}, \quad \: \text{since} \: D_{mk} = \mathbf{0}_{N}, \forall m \notin \mathcal{M}_{k}, \\
			&= \sqrt{\rho_{u}}\sum_{m = 1}^{M} \sum_{i=1}^{K}\hat{\vec g}_{mk}^H \vec D_{mk}\mathbf{A}_{m} \vec g_{mi} \sqrt{\eta_{i}} s_{i} + \sum_{m = 1}^{M} \hat{\vec g}_{mk}^H \vec D_{mk}\mathbf{A}_{m}\vec w_{m} + \sum_{m = 1}^{M} \hat{\vec g}_{mk}^H \vec D_{mk}\vec w_{m}^{q}.
		\end{aligned}
	\end{equation}
	\subsection{Achievable Uplink Rate}\label{sec:UCCF_Rate}
	In this subsection, the uplink rate performance of the UC CF-mMIMO system introduced is studied.  Specifically, using the ``use-and-then-forget" methodology popularized in \cite{marzetta2016fundamentals} and used in \cite{Ngo2017:CellFree}, we derive a lower bound for the achievable uplink rate of the $ k $th user in closed-form. It is assumed that only the knowledge of channel statistics is available at the CPU. The received signal in (\ref{Eq:UCCF_sk}) can be rearranged as 
	\begin{equation}\label{Eq:UCCF_skext}
		\begin{aligned}
			\hspace{-20mm}	\hat{s_k} &=  \underbrace{\sqrt{\rho_{u}\eta_{k}}\mathbb{E} \left[ \sum_{m=1}^{M} \hat{\vec g}_{mk}^H \vec D_{mk}\mathbf{A}_{m} \vec g_{mk}s_{k} \right]}_{\mathrm{DS}_{k}} +  \underbrace{\sqrt{\rho_{u} \eta_{k}}\left(\sum_{m=1}^{M} \hat{\vec g}_{mk}^H \vec D_{mk}\mathbf{A}_{m} \vec g_{mk}  - \mathbb{E} \left[ \sum_{m=1}^{M} \hat{\vec g}_{mk}^H \vec D_{mk}\mathbf{A}_{m} \vec g_{mk} \right] \right)s_{k}}_{\mathrm{BU}_{k}} \\	&+ \underbrace{\sqrt{\rho_{u}}\sum_{i \ne k}^{K} \sum_{m=1}^M \sqrt{\eta_{i}}\hat{\vec g}_{mk}^H \vec D_{mk}\mathbf{A}_{m} \vec g_{mi}s_{i}}_{\mathrm{IUI}_{k}} + \underbrace{\sum_{m=1}^M \hat{\vec g}_{mk}^H \vec D_{mk}\mathbf{A}_{m}\vec w_{m}}_{\mathrm{GN}_{k}} + \underbrace{\sum_{m=1}^M \hat{\vec g}_{mk}^H \vec D_{mk}\vec w_{m}^{q}}_{\mathrm{QN}_{k}},
		\end{aligned}
	\end{equation}
	where $\mathrm{DS}_{k}$, $\mathrm{BU}_{k}$, $\mathrm{IUI}_{k}$, $\mathrm{GN}_{k}$, and $\mathrm{QN}_{k}$ represents the desired signal, the beamforming gain uncertainty, the inter-user interference, the additive white Gaussian noise and the quantization noise, respectively. In (\ref{Eq:UCCF_skext}), it can be shown that the desired signal term is uncorrelated with the other terms since symbols for different users are independent, and noise is independent of the signal. Then, using the fact that uncorrelated Gaussian noise gives a lower bound on the capacity \cite{marzetta2016fundamentals,Ngo2017:CellFree}, the achievable uplink rate expression
	of the $ k $th user is given in the following theorem.
	\begin{theorem}\label{Thm:UCCF_uplinkrate}
		The achievable uplink rate for the $ k $th user in a UC CF-mMIMO system with MRC and a mixed ADC resolution profile at the APs is given by,
		\begin{equation}\label{Eq:UCCF_UPRatek}
			\begin{aligned}
				R_{k}^{UL} = \log_{2}\left(1 + \frac{\Gamma_{k}}{\Lambda_{k}} \right),
			\end{aligned}
		\end{equation}
		where
		\begin{equation}
			\Gamma_{k} = \rho_{u}\eta_{k} \left( \sum_{m=1}^{M} \gamma_{mk} \operatorname{Tr}\left( \vec D_{mk} \vec A_m\right)\right)^{2},
		\end{equation}
		and
		\begin{equation}\label{Eq:UCCF_Lambdak}
			\begin{aligned}
				\Lambda_{k} &= \rho_{u}\sum_{i \ne k}^{K}{\eta_i} \left( \sum_{m=1}^M    \gamma_{mk}   \frac{\beta_{mi}}{\beta_{mk}} \operatorname{Tr}\left(\vec D_{mk}\vec A_{m}\right)\right)^{2} \vert \boldsymbol{\phi}_{k}^{H}\boldsymbol{\phi}_{i} \vert^{2} \\&\qquad+\rho_{u} \sum_{i=1}^{K}\eta_{i} \sum_{m=1}^{M}\gamma_{mk}\beta_{mi} \operatorname{Tr}\left(\vec D_{mk} \mathbf{A}_{m}\right) + \sum_{m=1}^{M}\gamma_{mk}\operatorname{Tr}\left(\vec D_{mk} \mathbf{A}_{m}\right)\\
				&\qquad+\rho_{u}\sum_{m=1}^{M}\gamma_{mk}\left(\sum_{i=1}^{K}\eta_{i}  \gamma_{mi} \vert \boldsymbol{\phi}_{k}^{H}\boldsymbol{\phi}_{i} \vert^{2} \right)\operatorname{Tr}\left(\vec D_{mk} \mathbf{A}_{m} \left(\mathbf{I} - \mathbf{A}_{m}\right)\right). 
			\end{aligned}
		\end{equation}
	\end{theorem}
	\begin{proof}
		Please refer to Appendix \ref{Proof:UCCF_uplinkrate} for the proof.
	\end{proof}
	\section{Optimization Problem}\label{sec:UCCF_OptProb}
	This section first introduces the energy consumption model for the APs with mixed ADC resolution profiles. We then formulate the problem to optimize the power coefficient $ \eta_{k} $ and the antenna selection coefficient  $ d_{mk}^{n}$ jointly for maximizing the sum-rate under energy consumption constraints and present a tractable solution for the optimization problem.
	\subsection{Energy Consumption Model}
	We consider an energy consumption model similar to\cite{MixedADCCellFreeRician}.
	The overall power consumption of the $m$th AP is modeled as
	\begin{equation}\label{Eq:UCCF_EnergyConsumption}
		\begin{aligned}
			E_{AP}^{m}	&= c_0\sum_{n=1}^{N_1}d_{m}^{n} 2^{n} + 0.002 c_{1} + c_{2}\left(\sum_{n=N_{1}+1}^{N}d_{m}^{n}\right) ,
		\end{aligned}
	\end{equation} 
	where $ c_{0} = 3\times 10^{-5}$ is a constant that depends on the specific design of the ADC, $c_{1}$ is an indicator-variable related to $ d_{m}^{n} $. If $ d_{m}^{n} = 0 \ \forall n = 2,\dots,N_{1} $ then, $ c_{1} = 0 $; otherwise $ c_{1} = 1 $. Also, $c_{2} = 0.1229$ Watt represents the power consumption of the high-bit resolution ADCs. Furthermore, $d_{m}^{n}$ is an indicator-variable to denote the state of the $n$th antenna of the $m$th AP. It takes values $1$ or $0$ according to the following rule,
	\begin{equation}\label{Eq:UCCF_ASC}
		d_{m}^{n} = \begin{cases} 
			0 \quad \text{if} \quad d_{mk}^{n} = 0 \quad \forall \ k, \\
			1 \quad \text{otherwise}.
		\end{cases}
	\end{equation}
	The total power consumption of the network is given by $ E_{sys} = \sum_{m=1}^{M} E_{AP}^{m} $ and the maximum power consumption is represented by $E_{sys}^{max}$. Note that $E_{sys}^{max}$ corresponds to the value of $E_{sys}$ evaluated for $d_{m}^n=1$ for all $m,n$. 
	\subsection{Optimization Problem}
	We formulate the optimization problem with the sum-rate objective and constraints on the power control coefficient and energy consumption. The mathematical formulation of the optimization problem is as follows,
	\begin{equation}\label{Eq:UCCF_SumRate_Opt}
		\begin{aligned}
			&P1: \max_{\lbrace d_{mk}^{n} \rbrace, \boldsymbol{\eta}} \quad \sum_{k=1}^K R_{k}^{UL} \\
			\qquad \quad \  s.t. \ &  \  \quad 0 \le \eta_{k} \le 1, \quad \forall \ k = 1,\dots,K \\
			\qquad \quad \quad  \  \ &   \ \quad \sum_{m=1}^{M}\left[c_0\sum_{n=1}^{N_1}d_{m}^n2^{b_m^n} +c_1 + \sum_{n=N_{1}+1}^{N}d_{m}^nc_2\right] < E^{max},
		\end{aligned}
	\end{equation}
	where $E^{max}$ denotes the maximum energy consumption acceptable for the network. Note that the objective is a non-convex function, and the energy consumption constraint is a function of antenna selection coefficient $d_{mk}^{n}$ that takes binary values, either $0$ or $1$.
	\par The energy constraints and the use of mixed ADC architecture at the APs make our problem different from existing works like \cite{buzzi2019user}. Although employing an antenna with higher resolution ADCs reduces the quantization error and improves the rate, higher power dissipation at the ADC is also incurred, and the energy constraint cannot be met. Therefore, determining the suitable trade-off between the rate and energy consumption at the APs is essential. Hence, the antenna selection coefficients at the APs must be optimized to constrain the energy consumption at the APs and maximize the sum-rate.
	\par Note that (\ref{Eq:UCCF_SumRate_Opt}) is a joint optimization problem over the variables $\lbrace d_{mk}^{n} \rbrace$ and $\boldsymbol{\eta}$. Observe that $\eta_{k}$ are continuous variables, whereas $d_{mk}^{n}$ are discrete variables. Obtaining an optimal solution to the above problem is complicated due to the presence of integer antenna selection constraints. Also, the objective function for the $P1$ is not convex with respect to the power and the antenna-selection coefficients. Therefore, we propose a centrally implemented optimization technique that alternates between two algorithms sequentially. One algorithm is used to optimize over the discrete variable $d_{mk}^{n}$ for fixed $\boldsymbol{\eta}$, and another algorithm is used to optimize over the continuous variable $\boldsymbol{\eta}$ for fixed $d_{mk}^{n}$. 
	\subsubsection{Optimizing over $\boldsymbol{\eta}$}
	\par For a fixed $\lbrace d_{mk}^{n} \rbrace$, the optimization problem in (\ref{Eq:UCCF_SumRate_Opt}) reduces to
	\begin{equation}\label{Eq:UCCF_SumRate_OptOnlyeta}
		\begin{aligned}
			&P2: \max_{\boldsymbol{\eta}} \sum_{k=1}^K R_{k}^{UL} \\
			& \quad \quad \  s.t. \ \   0 \le \eta_{k} \le 1, \quad \forall \ k = 1,\dots,K.
		\end{aligned}
	\end{equation}
	For the sum-rate optimization in (\ref{Eq:UCCF_SumRate_OptOnlyeta}), the objective function is non-convex. Therefore, we adopt the SCA to solve it through a sequence of relaxed convex sub-problems. Using the rate expression derived in Theorem \ref{Thm:UCCF_uplinkrate}, the sum-rate objective function is transformed into a ratio of posynomials as follows
	\begin{equation}
		\begin{aligned}
			& \max_{\boldsymbol{\eta}} \sum_{k=1}^K \log_{2}\left(1+\frac{\Gamma_{k}}{\Lambda_{k}}\right) \\
			\Leftrightarrow \quad &\max_{\boldsymbol{\eta}} \log_{2} \left(\prod_{k=1}^{K} \left( 1+ \frac{\Gamma_{k}}{\Lambda_{k}} \right) \right) \\
			\stackrel{\left(a\right)}{\Leftrightarrow} \quad &\max_{\boldsymbol{\eta}} \prod_{k=1}^{K} \left( 1+\frac{\Gamma_{k}}{\Lambda_{k}} \right) \\ 
			\Leftrightarrow \quad &\min_{\boldsymbol{\eta}} \prod_{k=1}^{K} \left(\frac{\Lambda_{k}}{ \Lambda_{k} + \Gamma_{k} }\right), 
		\end{aligned}
	\end{equation}
	where $\left(a\right)$ follows from the fact that $\log$ is a monotonically increasing function.  Rearranging the terms in (\ref{Eq:UCCF_Lambdak}), we have
	\begin{equation}
		\begin{aligned}
			\Lambda_{k} &=  \sum_{i=1}^{K}\left( \delta_{k,i}^{\left(1\right)} + \delta_{k,i}^{\left(2\right)} \right)\eta_{i} + \sum_{i \ne k}^{K}  \delta_{k,i}^{\left(3\right)} \eta_{i} + \lambda_{k}^{\left(1\right)},
		\end{aligned}
	\end{equation}
	where
	\begin{equation}
		\begin{aligned}
			\delta_{k,i}^{\left(1\right)} &\triangleq \rho_{u}\sum_{m=1}^{M}\gamma_{mk}\beta_{mi} \operatorname{Tr}\left(\vec D_{mk} \mathbf{A}_{m}\right)
		\end{aligned}
	\end{equation}
	\begin{equation}
		\begin{aligned}
			\delta_{k,i}^{\left(2\right)} &\triangleq    \rho_{u} \sum_{m=1}^{M}\gamma_{mk}  \gamma_{mi}  \operatorname{Tr}\left(\vec D_{mk} \mathbf{A}_{m} \left(\mathbf{I} - \mathbf{A}_{m}\right)\right)\vert \boldsymbol{\phi}_{k}^{H}\boldsymbol{\phi}_{i} \vert^{2} 
		\end{aligned}
	\end{equation}
	\begin{equation}
		\begin{aligned}
			\delta_{k,i}^{\left(3\right)} &\triangleq       \rho_{u}\left( \sum_{m=1}^M    \gamma_{mk}   \frac{\beta_{mi}}{\beta_{mk}} \operatorname{Tr}\left(\vec D_{mk}\vec A_{m}\right)\right)^{2} \vert \boldsymbol{\phi}_{k}^{H}\boldsymbol{\phi}_{i} \vert^{2}
		\end{aligned}
	\end{equation}
	\begin{equation}
		\begin{aligned}
			\lambda_{k}^{\left(1\right)} &\triangleq \sum_{m=1}^{M}\gamma_{mk}\operatorname{Tr}\left(\vec D_{mk} \mathbf{A}_{m}\right). 
		\end{aligned}
	\end{equation}
	Similarly, 
	\begin{equation}
		\begin{aligned}
			\Lambda_{k} + \Gamma_{k} &= \sum_{i=1}^{K}\left( \delta_{k,i}^{\left(1\right)} + \delta_{k,i}^{\left(2\right)} + \delta_{k,i}^{\left(3\right)}\right)\eta_{i} + \lambda_{k}^{\left(1\right)}.
		\end{aligned}
	\end{equation}
	Note that, for a given $\lbrace d_{mk}^{n} \rbrace$, the $ \delta_{k,i}^{\left(1\right)} ,\delta_{k,i}^{\left(2\right)},\delta_{k,i}^{\left(3\right)}$ and $\gamma_{k}^{\left(1\right)}$ are dependent on only the system parameters which are known apriori and need not be optimized. Therefore, the optimization problem $P2$ is of the form
	\begin{equation}\label{Eq:UCCF_SumRate_OptOnlyetaRe}
		\begin{aligned}
			&P3:  \min_{\boldsymbol{\eta}} \prod_{k=1}^{K} \left(\frac{\sum_{i=1}^{K}\left( \delta_{k,i}^{\left(1\right)} + \delta_{k,i}^{\left(2\right)} \right)\eta_{i} + \sum_{i \ne k}^{K}  \delta_{k,i}^{\left(3\right)} \eta_{i} + \lambda_{k}^{\left(1\right)}}{ \sum_{i=1}^{K}\left( \delta_{k,i}^{\left(1\right)} + \delta_{k,i}^{\left(2\right)} + \delta_{k,i}^{\left(3\right)}\right)\eta_{i} + \lambda_{k}^{\left(1\right)} }\right) \\
			& \quad \quad \  s.t. \ \   0 \le \eta_{k} \le 1, \quad \forall \ k = 1,\dots,K.
		\end{aligned}
	\end{equation}
	The above objective function, a ratio of two posynomials, is a complementary GP, an intractable NP-hard problem \cite{Chiang2007GP}. In such a scenario, we can use the single condensation method, where the ratio is upper bounded by another posynomial \cite{TWRRelay}. Assume a function $ f\left(z\right) = \dfrac{h\left(z\right)}{g\left(z\right)} $, where both $ h\left(z\right) $ and $ g\left(z\right) $ are posynomials. The denominator term $ g\left(z\right) $ is lower-bounded by a monomial using the arithmetic-mean geometric-mean inequality. If $ z^{\left(i-1\right)} $ is the value of $ z $ at the iteration $ i - 1 $ of the SCA and $ g\left(z\right) \triangleq \sum_{k = 1}^{K}\mu_{k}\left(z\right) $, where $ \mu_{k}\left(z\right) $ are monomials, then
	\begin{equation}\label{AMGM}
		g\left(z\right) \ge \tilde{g}\left(z\right) = \prod_{k=1}^{K} \left(\dfrac{\mu_{k}\left(z\right)}{\nu_{k}\left(z^{\left(i-1\right)}\right)}\right)^{\nu_{k}\left(z^{\left(i-1\right)}\right)}
	\end{equation} 
	where $ \nu_{k}\left(z^{\left(i-1\right)}\right) = \dfrac{\mu_{k}\left(z^{\left(i-1\right)}\right)}{g\left(z^{\left(i-1\right)}\right)} = \dfrac{\mu_{k}\left(z^{\left(i-1\right)}\right)}{\sum_{k = 1}^{K}\mu_{k}\left(z^{\left(i-1\right)}\right)} $.
	Hence, the ratio of posynomials $ f\left(z\right) $ will be replaced by $ \tilde{f}\left(z\right) = \dfrac{h\left(z\right)}{\tilde{g}\left(z\right)} $, such that $ \tilde{f}\left(z\right) < f\left(z\right)$.
	In $P3$, $ g \left( \boldsymbol{\eta} \right) = \sum_{i = 1}^{K}\left(\delta_{k,i}^{\left(1\right)} + \delta_{k,i}^{\left(2\right)} + 
	\delta_{k,i}^{\left(3\right)} \right) \eta_{i}  + \lambda_{k}^{\left(1\right)} $ is a posynomial with $ K+1 $ terms. It can be converted into a monomial $ \tilde{g}\left(\boldsymbol{\eta} \right) $ by using the single condensation method. Finally, the SCA algorithm is utilized to solve $P3$, details of which are given in Algorithm \ref{Algo:UCCF_SCAAlgorithm}.  Here, OF stands for objective function in $P3$ and $ v $ is a parameter to control the accuracy of the algorithm. 
	
	\begin{algorithm}
		\caption{SCA Algorithm for (\ref{Eq:UCCF_SumRate_OptOnlyetaRe})}
		\begin{algorithmic}[1]\label{Algo:UCCF_SCAAlgorithm}
			\STATE \textit{\textbf{Initialization}}: Set $ i=1 $ and Select a feasible initial value of $ \boldsymbol{\eta} $
			\REPEAT
			\STATE Convert the problem in GP using single condensation method.
			\STATE Solve the relaxed convex sub-problem using interior-point method.
			\UNTIL $ \vert \text{OF}^{\left(i-1\right)} - \text{OF}^{i} \vert \le v$
		\end{algorithmic}
	\end{algorithm}
	\subsubsection{Optimizing over $d_{mk}^{n}$}
	Finally, to optimize over the $d_{mk}^{n}$, we use the BPSO, a meta-heuristic algorithm. A particle swarm heuristic based optimization problem changes the "trajectories'' of a population of "particles'' through the solution space of optimization problem. This change of "trajectory'' is done on the basis of each particle's previous best performance and the best performance of all particles. In binary particle swarm optimization (BPSO), the ``trajectories'' of a particle are modified in a probabilistic manner such that a coordinate will be assigned a zero or one value \cite{kennedy1997discrete}. Let $\boldsymbol{\epsilon}$ is a $MN \times K$ matrix with binary entries representing $ \lbrace d_{mk}^{n} \rbrace $. BPSO starts with generating an initial population of $ T $ particles, which in our case are the $T$ feasible solutions for the antenna selection coefficients ($d_{mk}^{n}$) represented by $\boldsymbol{\epsilon}^{\left( t \right)}, \ \forall t = 1,\cdots,T$. In each iteration, we first determine the optimal power allocation using SCA and then evaluate objective function (sum-rate in our case) for each of the particles $\boldsymbol{\epsilon}^{\left( t \right)}$. Each particle $\boldsymbol{\epsilon}^{\left( t \right)}$ maintains a record of the position of its previous best performance denoted by $ \boldsymbol{\epsilon}^{t,local}$ in terms of objective function (sum-rate) and the best performance of all particles denoted by $\boldsymbol{\epsilon}^{max}$. Velocity of each particle is updated based on $ \epsilon^{t,local} $ and $\epsilon^{max}$ according to (\ref{VelocityBPSO}). Next, a sigmoid transform is used to update the particles in a probabilistic manner given in (\ref{PositionUpdateBPSO}). The algorithm stops after either a predetermined number of iterations $I_{max}$ or convergence. The entire heuristic is provided in Algorithm \ref{Algo:UCCF_BPSOGP}.
	\begin{algorithm}
		\caption{BPSO with SCA Algorithm}
		\begin{algorithmic}[1]\label{Algo:UCCF_BPSOGP}
			\STATE \textit{\textbf{Initialization}}: Generate $ T $ particles $ \boldsymbol{\epsilon}^{\left(t\right)}, t = 1,\dots,T $ and set $ i = 1 $
			\REPEAT 
			\STATE Find $\boldsymbol{\eta}$ by solving the Algorithm \ref{Algo:UCCF_SCAAlgorithm} for each particle $ \boldsymbol{\epsilon}^{\left(t\right)} $.
			\STATE Compute the value of objective function \textit{i.e.}, sum-rate for each particle $ \boldsymbol{\epsilon}^{\left(t\right)} $, $ R^{(t)}(i) $
			\STATE Find $ \left(t_{m},i_{m}\right) = \arg \max\limits_{t,i} \ R^{(t)}(i)  $, Then set $ R_{max} = R^{(t_{m})}(i_{m}) $ and $ \boldsymbol{\epsilon}^{max} = \epsilon^{t_{m}}(i_{m}) $ 
			\STATE Get $ i_{t} = \arg\max\limits_{i} \ R^{(t)}(i) $ and set $ R^{t,local}  =R^{(t)}(i_{t}) \ \forall t $ and $ \boldsymbol{\epsilon}^{t,local}  = \epsilon^{(t)}(i_{t}) \ \forall t$ 
			\STATE Calculate velocity for each particle
			\begin{equation}\label{VelocityBPSO}
				V^{(t)}(i) = \Omega V^{(t)}(i-1) + \psi_{1}(i)\left(\boldsymbol{\epsilon}^{t,local}(i) - \boldsymbol{\epsilon}^{(t)}(i) \right) + \psi_{2}\left(\boldsymbol{\epsilon}^{max}(i) - \boldsymbol{\epsilon}^{(t)}(i) \right), 
			\end{equation}
			where $ \Omega = 0.9 - \frac{i\left(0.9 - 0.2\right)}{I_{max}} $ is the inertia weight and $ \psi_{1}, \psi_{2} \in \left[0,2\right] $ are two random positive numbers.
			\STATE Update the particle's position as follows
			\begin{equation}\label{PositionUpdateBPSO}
				\boldsymbol{\epsilon}^{(t)}(i+1) = \begin{cases*}
					1 \quad \text{if}  \ r_{rand} < \dfrac{1}{1 + e^{-V^{(t)}(i) }} \\
					0 \quad \text{otherwise}
				\end{cases*}
			\end{equation} 
			where $ r_{rand} $ is a random number generated from uniform distribution in $ \left[0,1\right] $
			\STATE increment $ i = i+1 $
			\UNTIL $ \vert \text{OF}^{\left(i-1\right)} - \text{OF}^{i} \vert \le v$
		\end{algorithmic}
	\end{algorithm}
	\section{Simulation Results}\label{sec:UCCF_Results}
	In this section, we study the performance of UC CF-mMIMO and compare the performance with CF-mMIMO systems. Most system parameters is similar to that used in \cite{Ngo2017:CellFree}, except the fact that we consider a mixed-ADC resolution. The $M$ APs and $K$ users are dispersed in a square of area $D \times D~ \text{km}^{2}$.  The large-scale fading coefficients, $\{\beta_{mk}\}$ modelling the path loss
	and shadow fading are selected as follows:
	\begin{equation}
		\beta_{mk}= PL_{mk}10^{\frac{\sigma_{th}z_{mk}}{10}}.
	\end{equation}
	Here, $PL_{mk}$ represents the path loss, $\sigma_{th}$ represents the standard deviation of the shadowing and $z_{mk} \sim \mathcal{N }(0, 1)$. The relation between the path loss $PL_{mk}$ and the distance $d_{mk}$ between the $m$-th AP and $k$-th user is obtained using the three slope model \cite[Eq. 52]{Ngo2017:CellFree}. The other parameters used in the simulation are summarized in Table \ref{tab:params}. 
	\begin{table}[h]
		\centering
		\begin{tabular}{|l|r|}
			\hline
			Parameter & value \\
			\hline
			\hline
			Carrier frequency & $1.9~GHz$ \\
			\hline
			Bandwidth & $20~MHz$\\
			\hline
			Noise figure & $9$ dB\\
			\hline
			AP antenna height & $15~m$\\
			\hline
			User antenna height & $1.65~m$\\
			\hline
			$\sigma_{sh}$ & $8$ dB\\
			\hline
			$\bar \rho_p$, $\bar \rho_u$ & $100~mW$\\
			\hline
			$\tau_{c}$, $\tau_{p}$ & $10$\\
			\hline
		\end{tabular}
		\caption{Simulation parameters}
		\label{tab:params}
	\end{table}
	The normalized transmit SNRs $ \rho_{p}$ and $ \rho_{u}$ are obtained by dividing the transmit powers $\bar \rho_{p}$ and $\bar \rho_{u}$ by the noise power, respectively.  Throughout the simulations, we have taken $N$, $N_1$, $N_2$ and $L$ to be $4$, $3$, $1$ and $5$, respectively, unless mentioned otherwise.
	\par Note that in CF-mMIMO, each user is served by all the APs. Hence, only the power coefficients are optimized according to the objective function, and all the antenna selection coefficients, \textit{i.e.}, $d_{mk}^{n}$ are set to be $1$. Therefore, to ensure a fair comparison between both systems, we plot the sum-rate energy efficiency (SREE), defined as the ratio of the sum-rate to the total energy consumption at the APs.
	\begin{figure}[h]
		\begin{subfigure}{0.48\textwidth}
			\includegraphics[width=\textwidth]{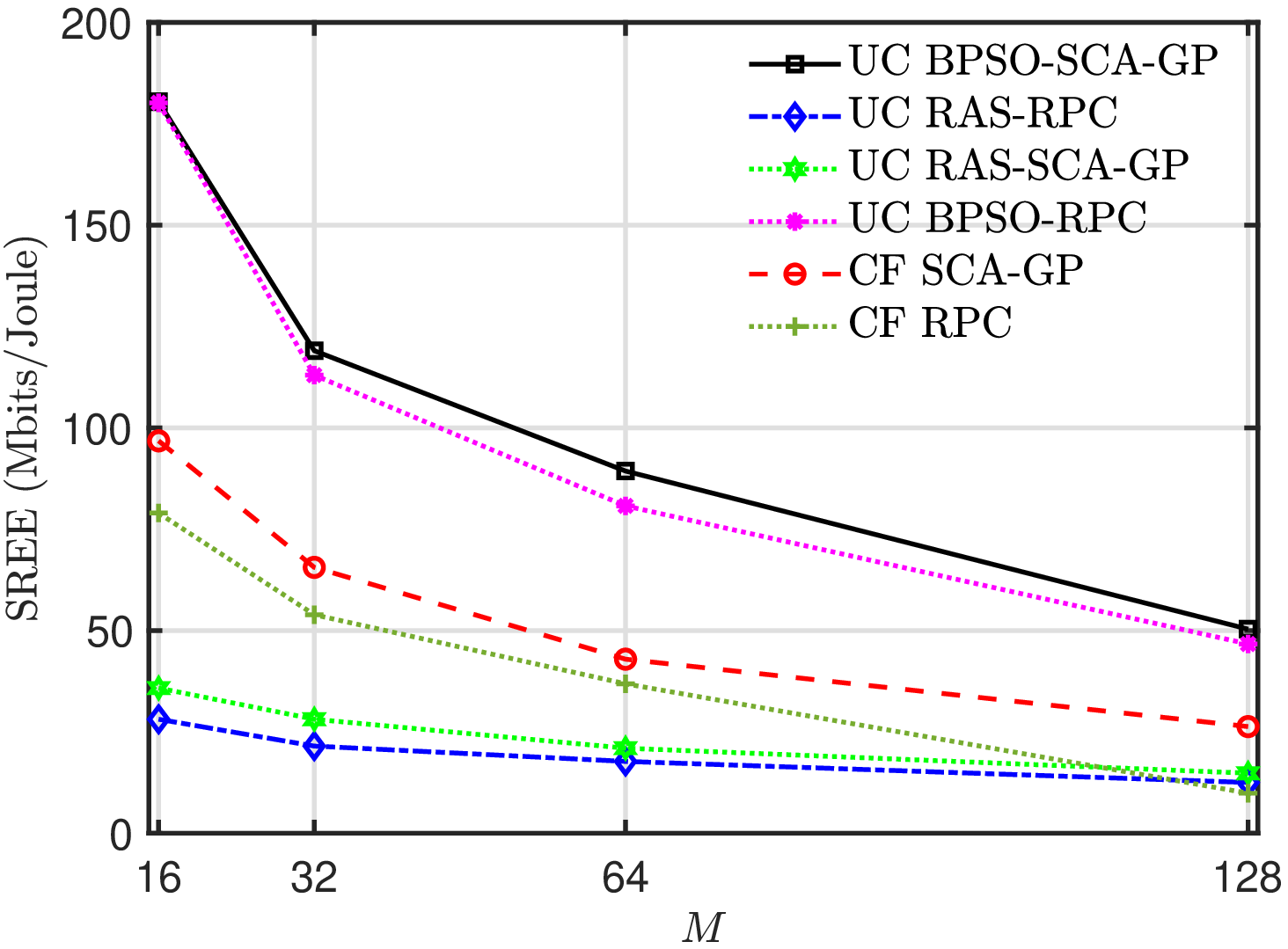}
			\caption{When $E^{max}=0.75 E_{sys}^{max}$}
			\label{Fig:SREE_M_Vary_K_8_EC_75}
		\end{subfigure}
		\hspace{3mm}
		\begin{subfigure}{0.48\textwidth}
			\includegraphics[width=\textwidth]{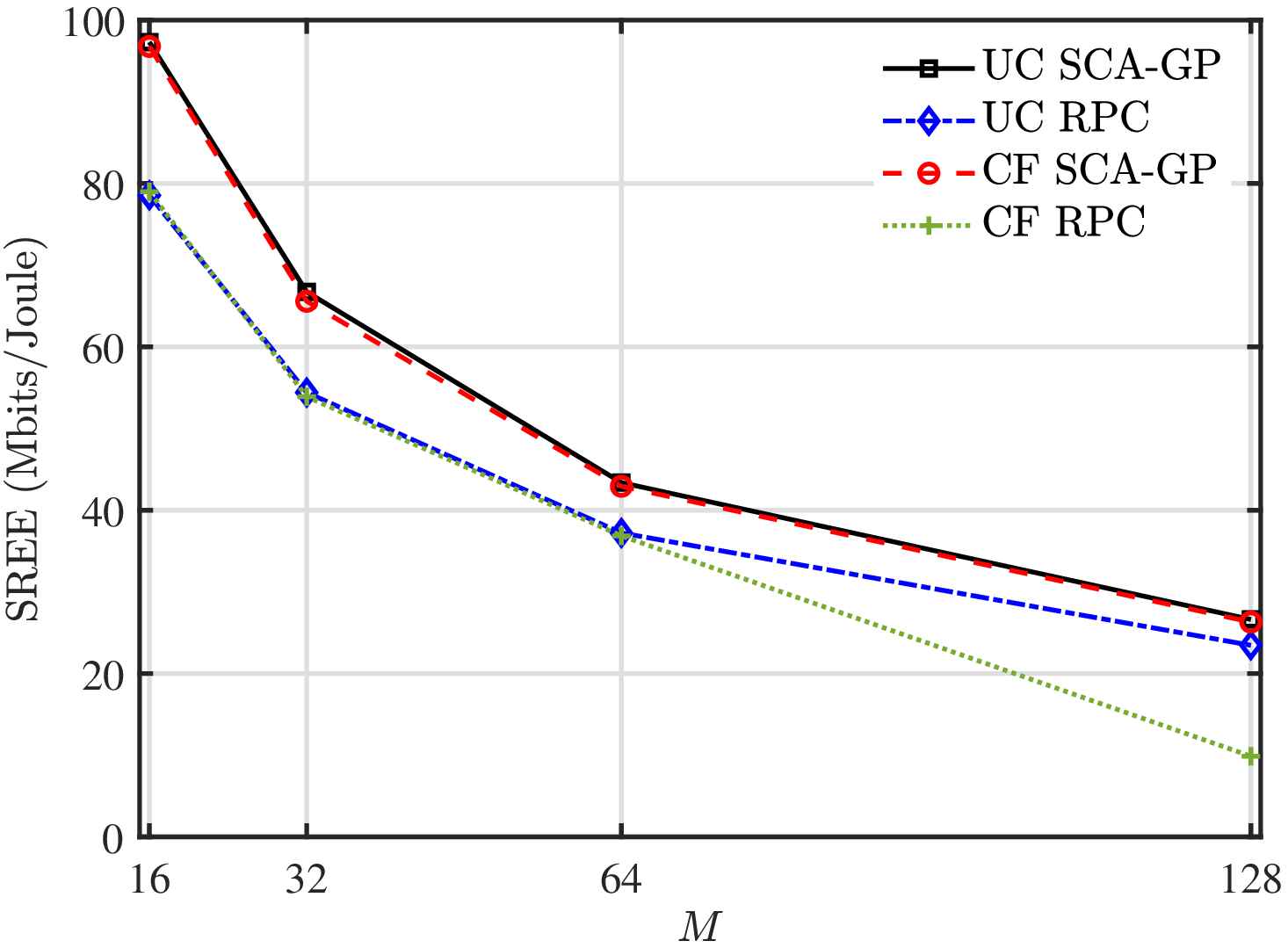}
			\caption{When $E^{max}=E_{sys}^{max}$}
			\label{Fig:SREE_M_Vary_K_8_EC_100}
		\end{subfigure}
		\caption{SREE versus $M$ for $K = 8, L=5, \kappa=0.25$}
		\label{figure1}
	\end{figure}
	\begin{figure}[h]
		\begin{subfigure}{0.48\textwidth}
			\includegraphics[width=\textwidth]{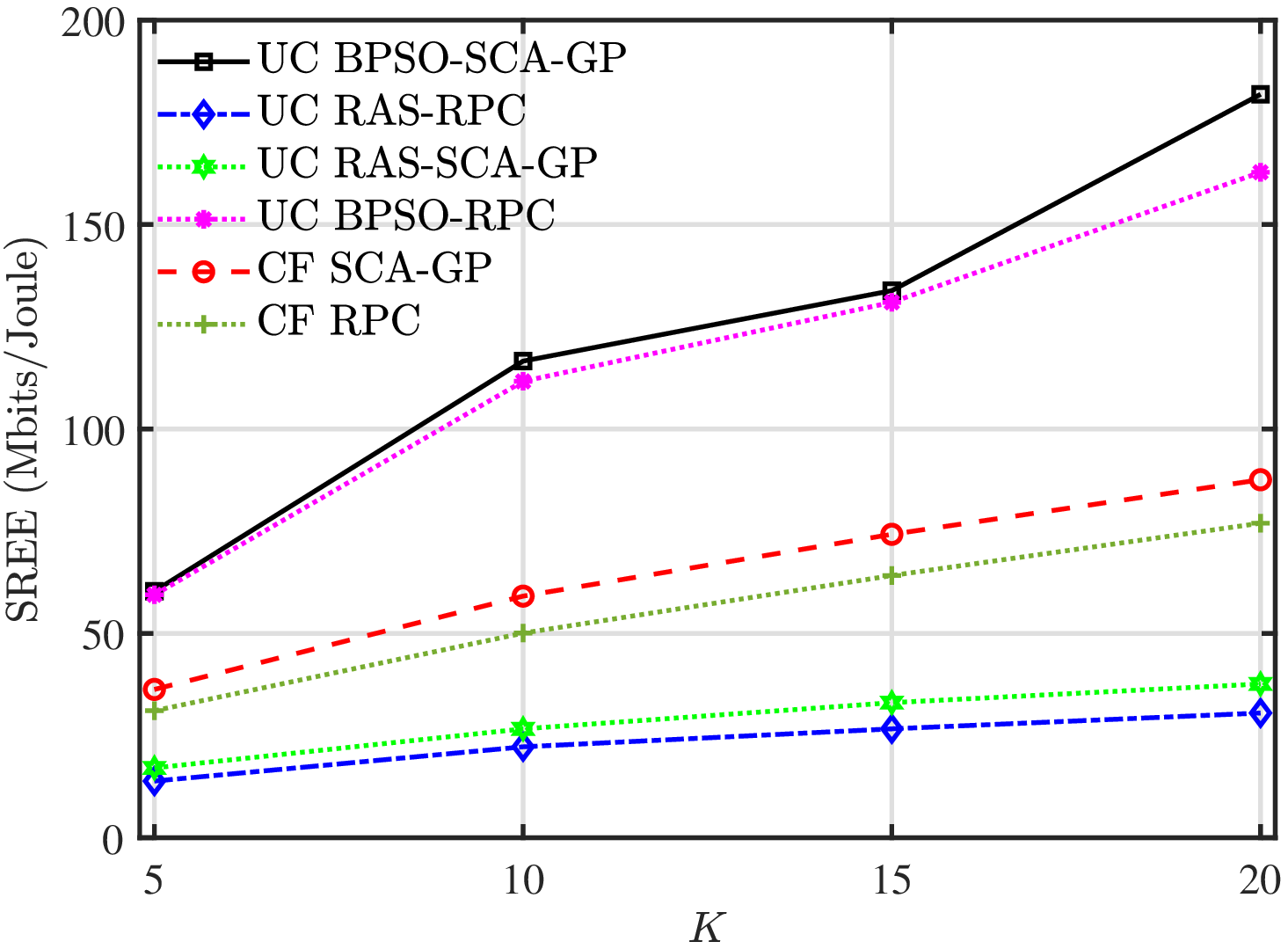}
			\caption{When $E^{max}=0.75 E_{sys}^{max}$}
			\label{Fig:SREE_K_Vary_M_50_EC_75}
		\end{subfigure}
		\hspace{3mm}
		\begin{subfigure}{0.48\textwidth}
			\includegraphics[width=\textwidth]{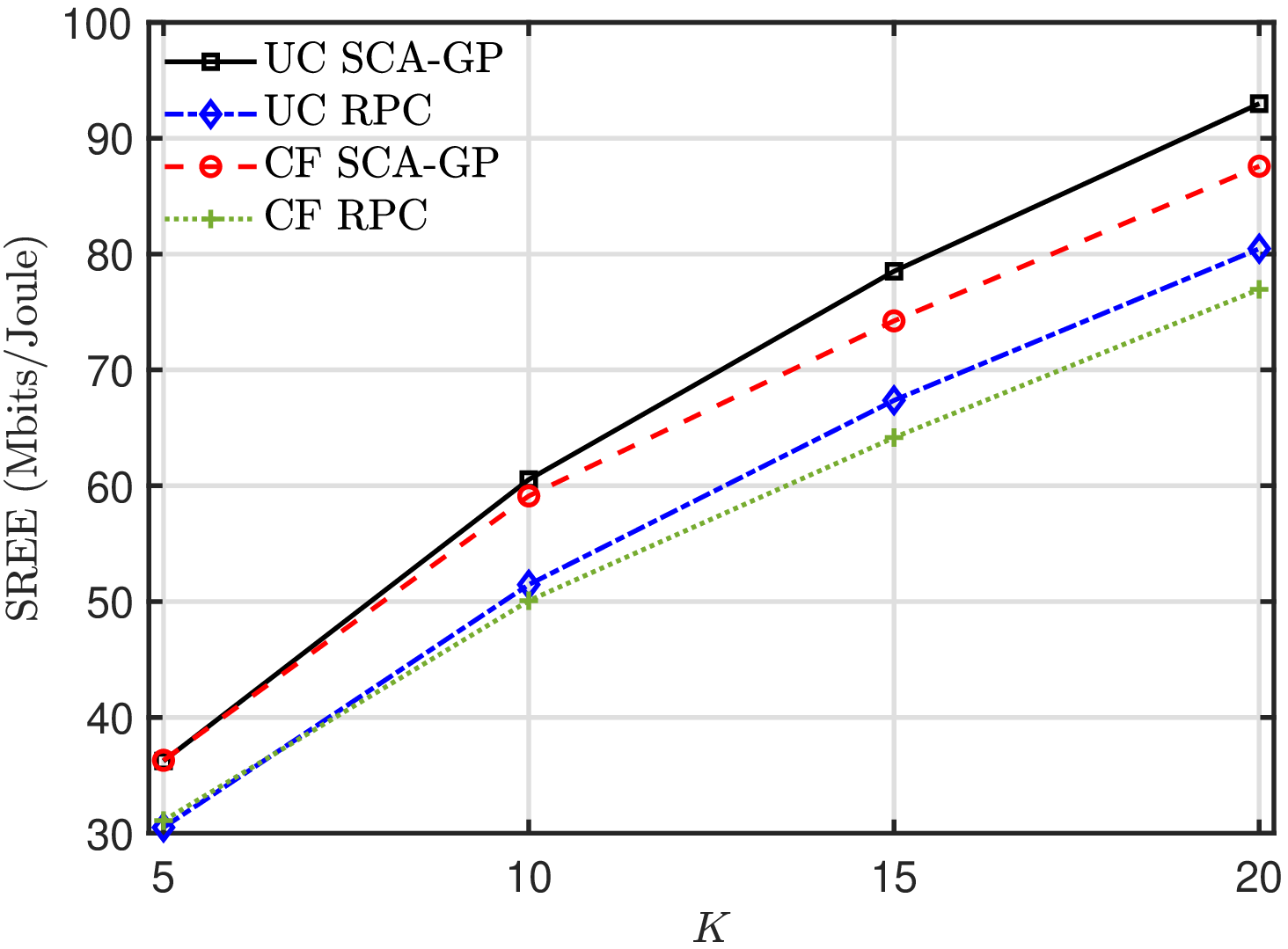}
			\caption{When $E^{max}=E_{sys}^{max}$}
			\label{Fig:SREE_K_Vary_M_50_EC_100}
		\end{subfigure}
		\caption{SREE versus $K$ for $M = 50, L=5, \kappa=0.25$}
		\label{figure2}
	\end{figure}
	Fig. \ref{figure1} and Fig. \ref{figure2} compare the SREE for UC CF-mMIMO, and CF-mMIMO for two different choices of $E^{max}$ for varying $M$ and $K$. The different schemes used for comparison are
	\begin{itemize}
		\item UC BPSO-SCA-GP: Joint Optimization of $\{d_{mk}^{n}\} $ and  $\vec{\eta}$ using Algorithm \ref{Algo:UCCF_BPSOGP}.
		\item UC random antenna selection (RAS) - random power coefficient (RPC): A random choice for $\{d_{mk}^{n}\} \in \{0,1\}$ and $\eta_{k} \sim \mathcal{U}\left[0,1\right]$.
		\item UC RAS-SCA-GP: A random choice for $\{d_{mk}^{n}\}$ and $\vec{\eta}$ is optimized via Algorithm \ref{Algo:UCCF_SCAAlgorithm}.
		\item UC BPSO-RPC: Algorithm \ref{Algo:UCCF_BPSOGP} with step $3$ replaced $\eta_{k} \sim \mathcal{U}\left[0,1\right]$.
		\item CF SCA-GP: All $d_{mn}^{k} = 1$ and $\vec{\eta}$ is optimized via Algorithm \ref{Algo:UCCF_SCAAlgorithm}.
		\item CF RPC: All $d_{mn}^{k} = 1$ and $\eta_{k} \sim \mathcal{U}\left[0,1\right]$.
	\end{itemize}
	It is evident from Fig. \ref{Fig:SREE_M_Vary_K_8_EC_75} and Fig. \ref{Fig:SREE_K_Vary_M_50_EC_75} that for the same system settings, the use of BPSO-SCA-GP demonstrates the best performance of all the schemes. Note that UC BPSO-SCA-GP performs better than CF SCA-GP because of the following two reasons: (i) In UC CF-mMIMO, each AP serves only the users with the best channel estimates. However, in a CF-mMIMO system, the AP serves users even with poor channel estimates, resulting in the overall performance degradation. (ii) The joint optimization of $\{d_{mk}^{n}\}$ and  $\vec{\eta}$ using Algorithm \ref{Algo:UCCF_BPSOGP} ensures that the sum-rate is maximized while satisfying the energy constraint (by selecting the appropriate antennas) resulting in a superior SREE. Note that the performance of UC BPSO-RPC is very close to the version obtained using UC BPSO-SCA-GP in both Fig. \ref{Fig:SREE_M_Vary_K_8_EC_75} and Fig. \ref{Fig:SREE_K_Vary_M_50_EC_75}. Thus it can be concluded that antenna selection is more crucial than optimizing the power coefficients. Moreover, replacing the SCA algorithm in BPSO-SCA-GP with random power allocation saves execution time. 
	\par From Fig. \ref{Fig:SREE_M_Vary_K_8_EC_75}, we can observe that SREE decreases as $M$ increases. As $M$ increases, energy consumption increases, but the sum-rate does not increase in the same proportion since $K$ is fixed. In contrast, in Fig. \ref{Fig:SREE_K_Vary_M_50_EC_75}, we can see that SREE increases as $K$ increases for a fixed $M$. Fig. \ref{Fig:SREE_M_Vary_K_8_EC_100} and Fig. \ref{Fig:SREE_K_Vary_M_50_EC_100} shows that in terms of SREE, for $100\%$ energy consumption, UC CF-mMIMO and CF-mMIMO have similar performances. Furthermore, note that randomized allocation by UC-RAS-RPC achieves the lowest SREE in all figures, reinforcing the importance of optimizing the system parameters by other schemes.
	\begin{figure}[h]
		\begin{subfigure}{0.48\textwidth}
			\includegraphics[width=\textwidth]{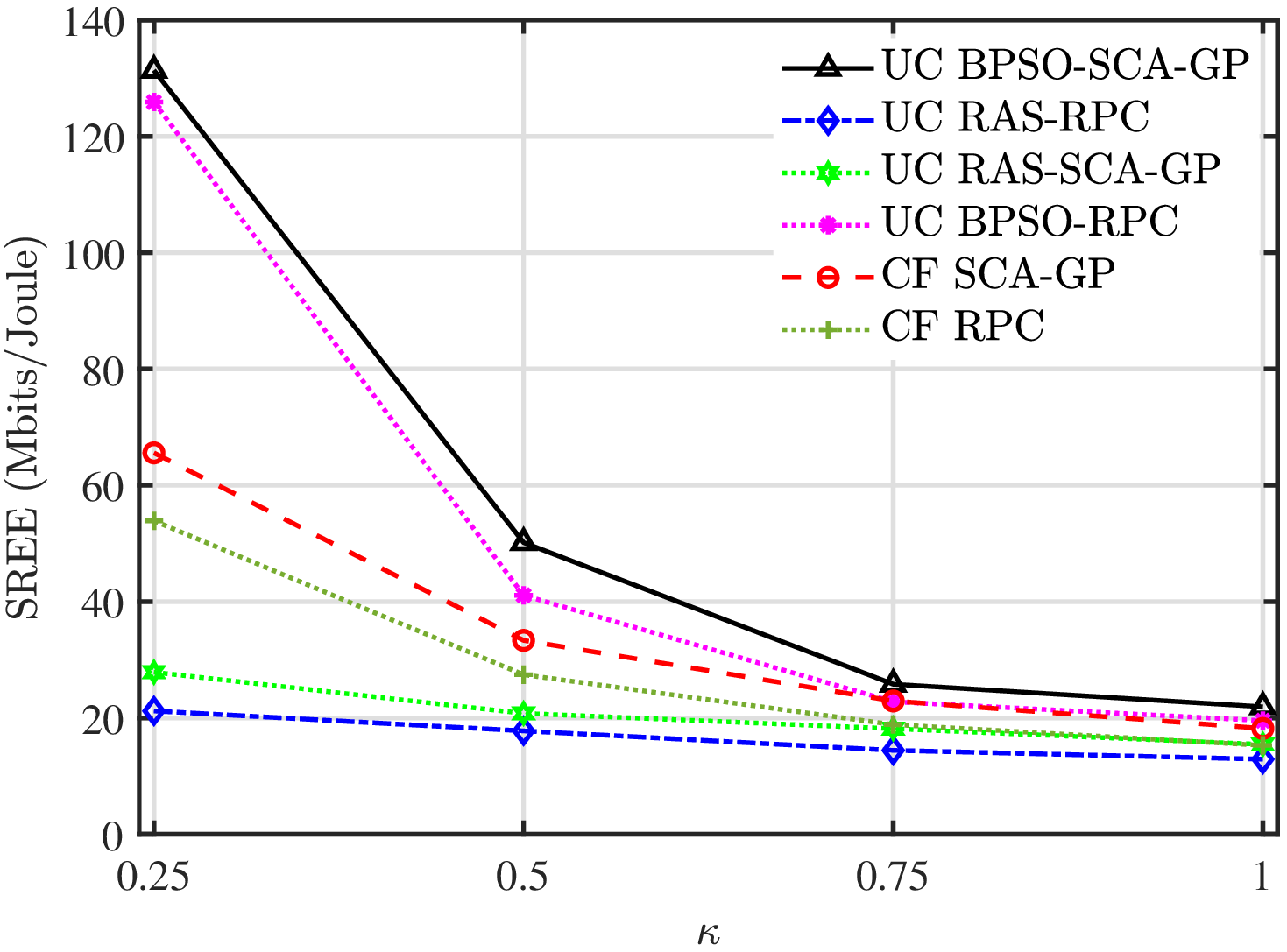}
			\caption{When $E^{max}=0.75 E_{sys}^{max}$}
			\label{Fig:SREE_Kappa_Vary_EC_75}
		\end{subfigure}
		\hspace{3mm}
		\begin{subfigure}{0.48\textwidth}
			\includegraphics[width=\textwidth]{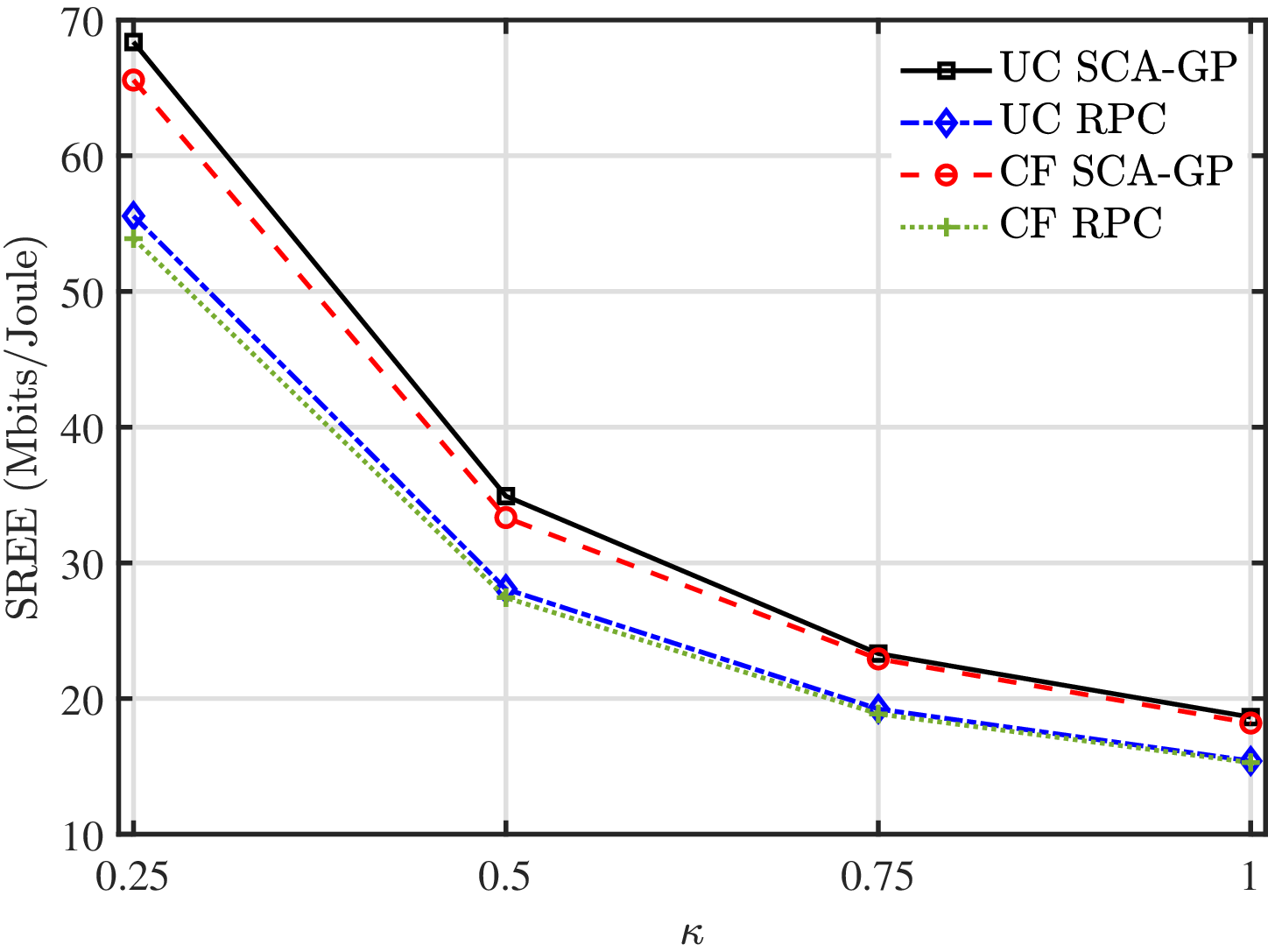}
			\caption{When $E^{max}=E_{sys}^{max}$}
			\label{Fig:SREE_Kappa_Vary_100}
		\end{subfigure}
		\caption{SREE versus $\kappa$ for $M = 32, K = 8, L=5$}
		\label{figure3}
	\end{figure}   
	\par In Fig. \ref{figure3}, we show the performance of our schemes for variations in the ratio of high-bit resolution ADCs to the total number of ADCs, denoted by $\kappa= \frac{N_2}{N}$. Here, we assume $N=4$ and $N_2$ will vary according to the chosen $\kappa$. Note that, for $\kappa=0.25$, i.e., for the case of only one high-bit resolution ADC, UC BPSO-SCA-GP performs significantly better than all the other schemes except UC-BPSO-RPC. This reiterates the idea that antenna selection is essential to maximize the SREE. With an increase in $\kappa$, the antennas are equipped with similar ADCs, the performance gap between the schemes decreases. Also, with a relaxation in the energy constraint in Fig. \ref{Fig:SREE_Kappa_Vary_100}, we observe a similar trend, but the SREE of UC CF-mMIMO decreases.  
	\begin{figure}[h]
		\begin{subfigure}{0.48\textwidth}
			\includegraphics[width=\textwidth]{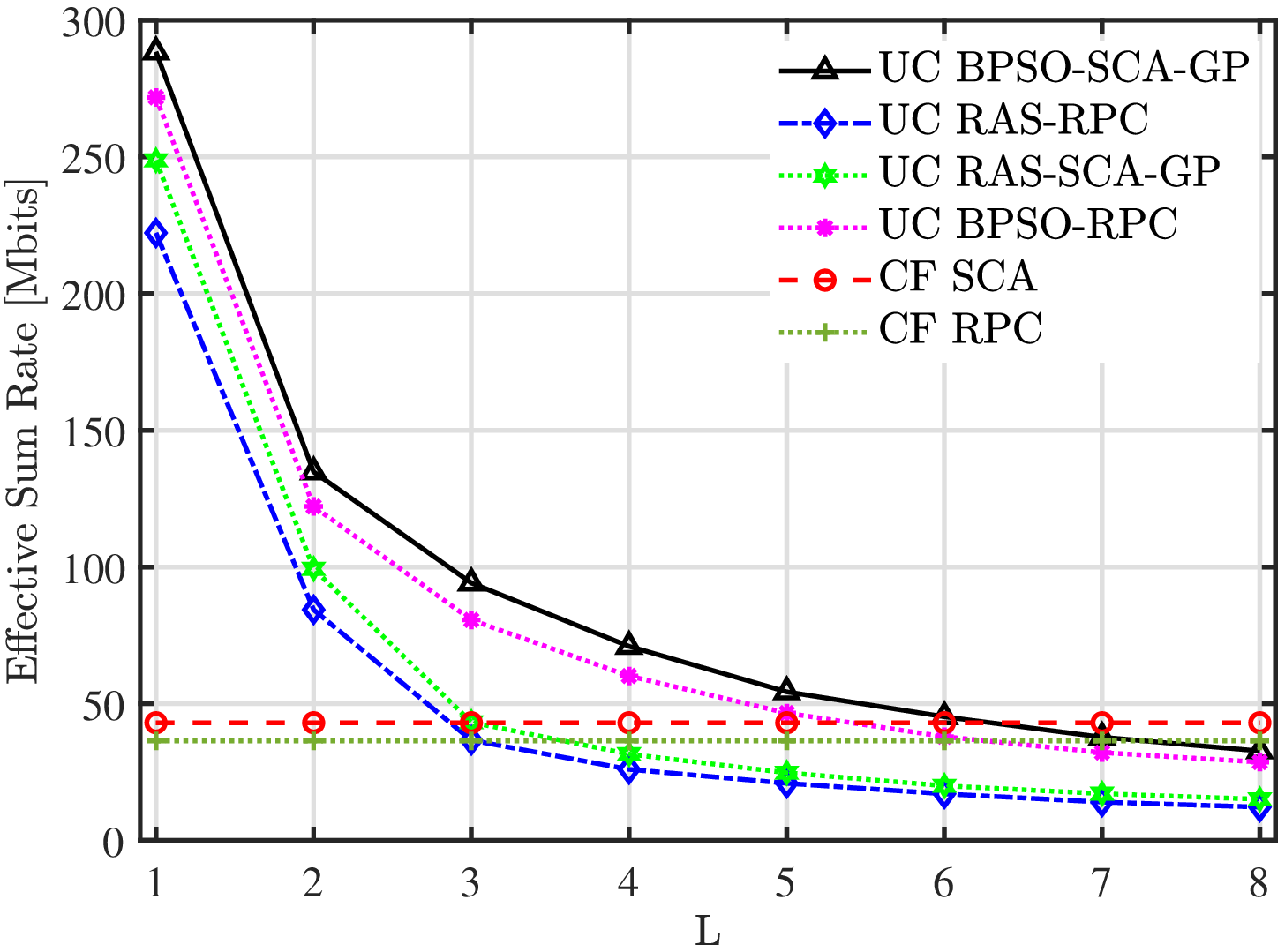}
			\caption{When $E^{max}=0.75 E_{sys}^{max}$}
			\label{Fig:SREE_L_Vary_EC_75}
		\end{subfigure}
		\hspace{3mm}
		\begin{subfigure}{0.48\textwidth}
			\includegraphics[width=\textwidth]{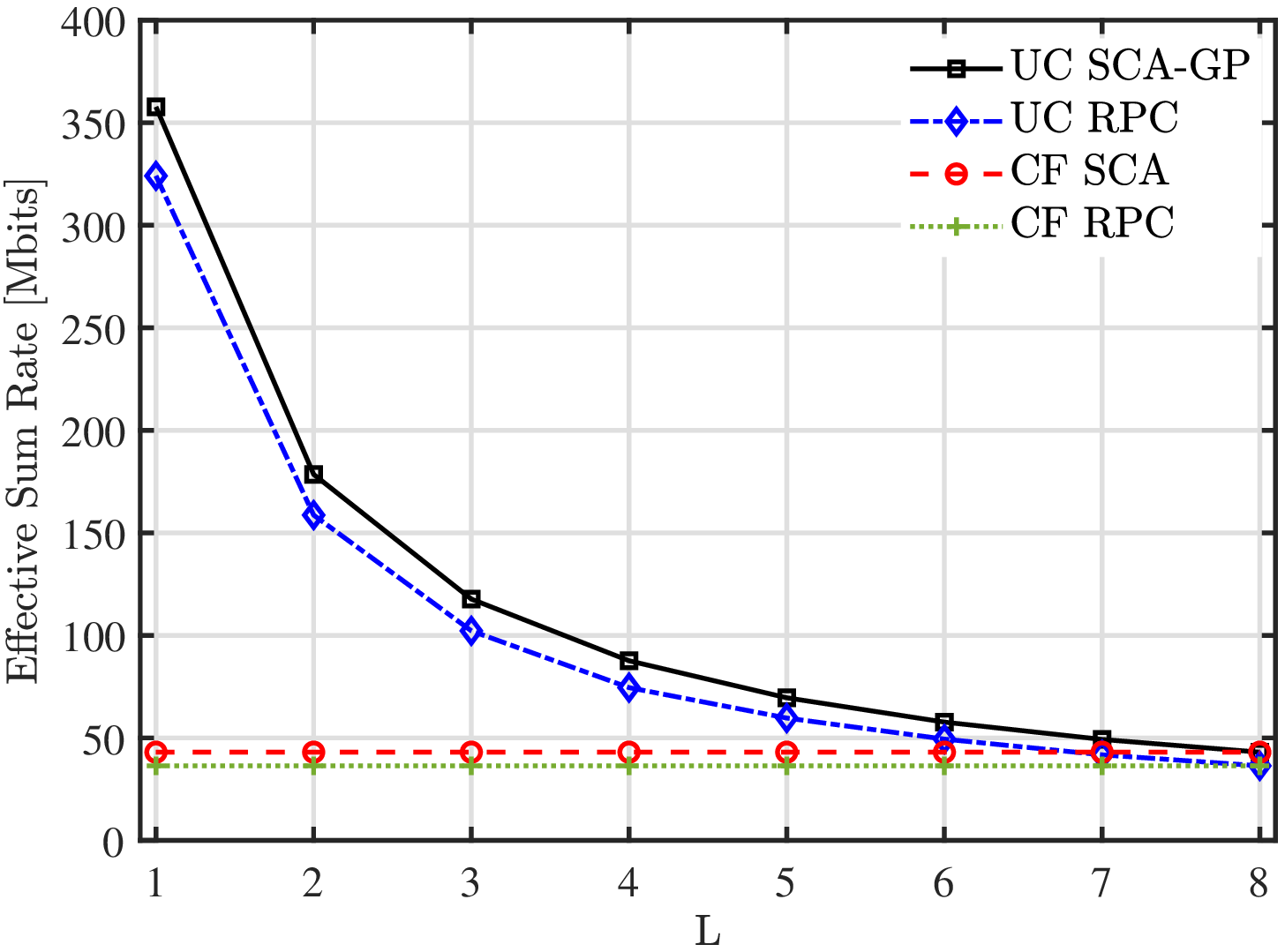}
			\caption{When $E^{max}=E_{sys}^{max}$}
			\label{Fig:SREE_L_Vary_100}
		\end{subfigure}
		\caption{SREE versus $L$ for $M = 64, K = 8, \kappa=0.25$}
		\label{figure4}
	\end{figure}  
	Finally, in Fig. \ref{figure4}, we plot the effective sum-rate, which is the sum-rate normalized by the number of users $L$ served per AP, for various values of $L$. We observe that with an increase in $L$, the interference increases at the APs, and hence, the effective sum-rate decreases. However, in contrast to SREE, for the case of effective sum-rate, the performance degradation from BPSO-SCA-GP to RAS-RPC is significantly lower. This is because BPSO-SCA-GP increases the sum-rate by reducing the interference via switching off antennas, thereby saving energy, and this advantage of BPSO-SCA-GP over RAS-RPC is reflected better in SREE of earlier figures.  
	\subsubsection*{Convergence \& Complexity} 
\begin{figure}[h]
		\begin{subfigure}{0.48\textwidth}
			\includegraphics[width=\textwidth]{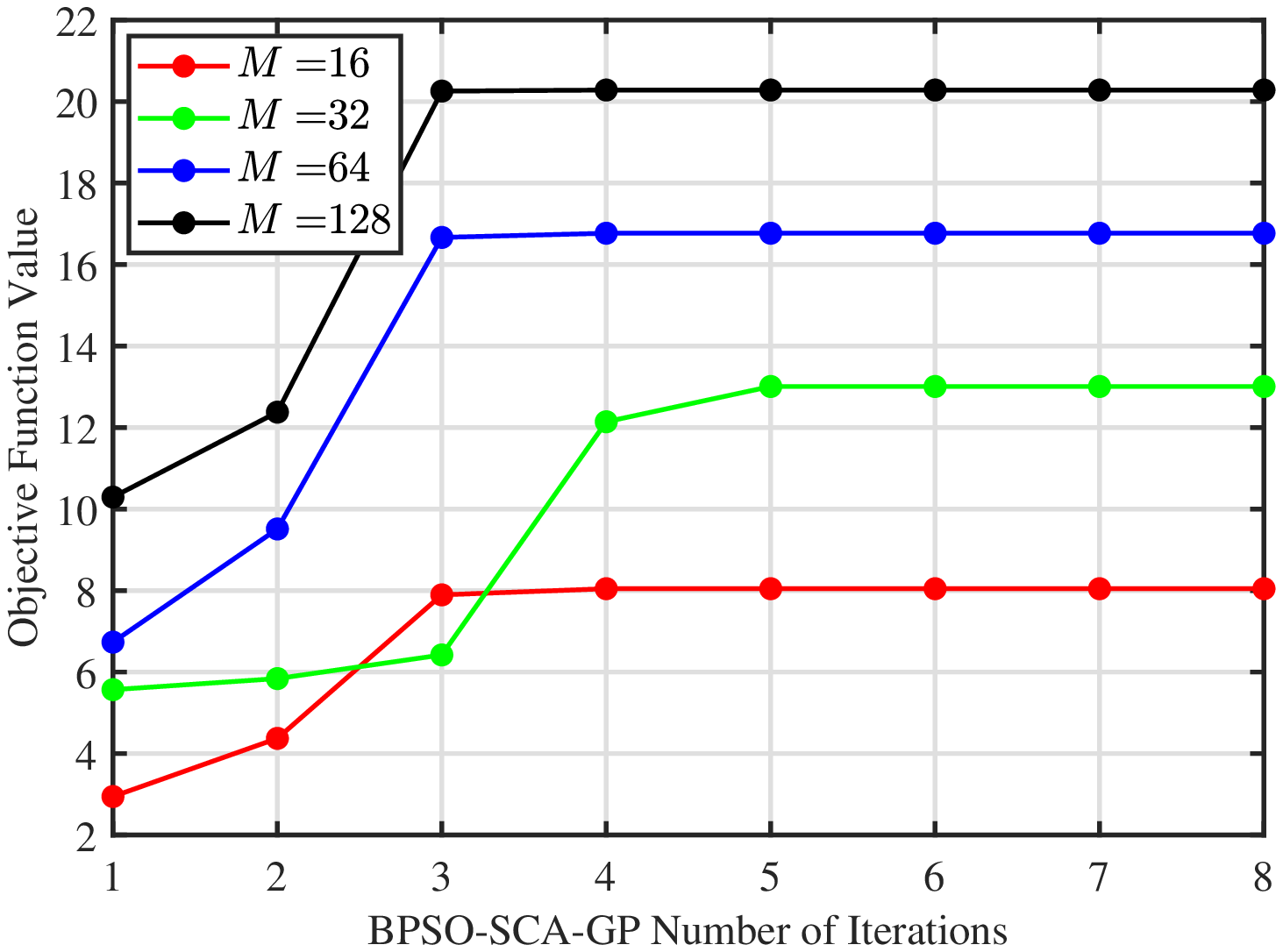}
			\caption{For various $M$ and $K=8$}
			\label{Fig:OF_Iteration_VaryM}
		\end{subfigure}
		\hspace{3mm}
		\begin{subfigure}{0.48\textwidth}
			\includegraphics[width=\textwidth]{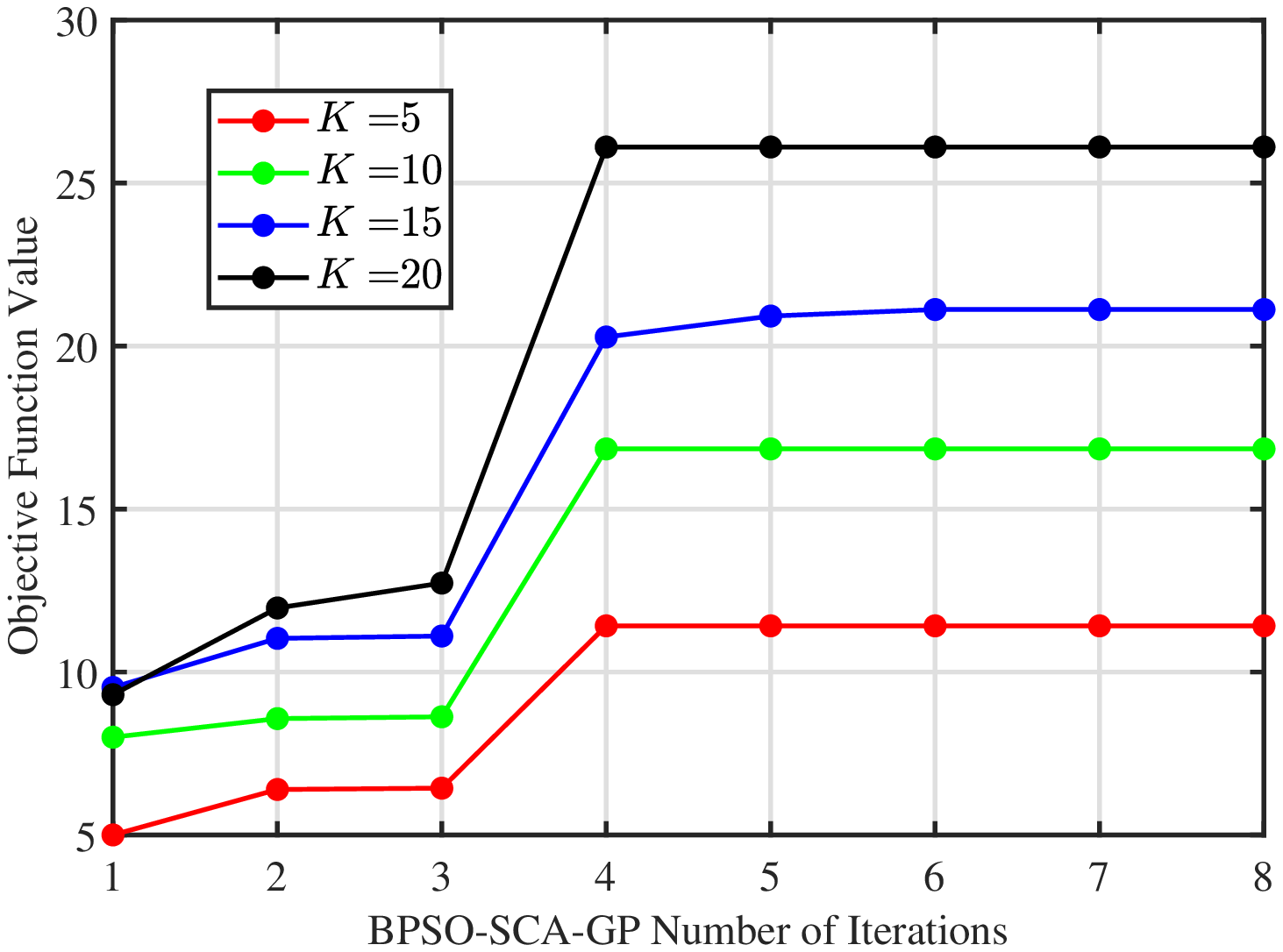}
			\caption{For various $K$ and $M=50$}
			\label{Fig:OF_Iteration_VaryK}
		\end{subfigure}
		\caption{Convergence of BPSO-SCA-GP}
		\label{figure_convergence}
	\end{figure}

Next, to study the convergence behavior of the system, we demonstrate the growth in the value of the objective function (sum-rate) for the increasing iteration number of our alternating optimization routine. Fig \ref{Fig:OF_Iteration_VaryM} and Fig \ref{Fig:OF_Iteration_VaryM} includes such plots for different values of $M$ and $K$ respectively. We can observe that for all choices of $M$ and $K$, the algorithm is monotonic, i.e., with every iteration, the values of the objective function increase and converges to a constant value. This indicates that each sub-problems maximizes the objective function and hence plays a role in moving the solution towards the optimum in very few iterations. 
	\par Note that one possible concern regarding the algorithm's performance can be the sub-optimality of the proposed solution due to the use of BPSO for optimizing the antenna selection. However, it will be a computationally intensive task to verify the closeness between the optimal solution obtained by a brute force search over all possible antenna selection choices and the solution proposed by BPSO. For example, consider a system with $M$ APs, each with $N$ antennas and  $K$ users. There are $2^{MNK}$ possible choices of antenna resolution configurations. Hence, an exhaustive search of all the possible combinations will be tedious even for moderate values of $M$, $N$, and $K$. Hence, we study the performance of the BPSO algorithm using a minimal system setting. 
	
\begin{figure}[h]
		\centering
		\includegraphics[scale=0.6]{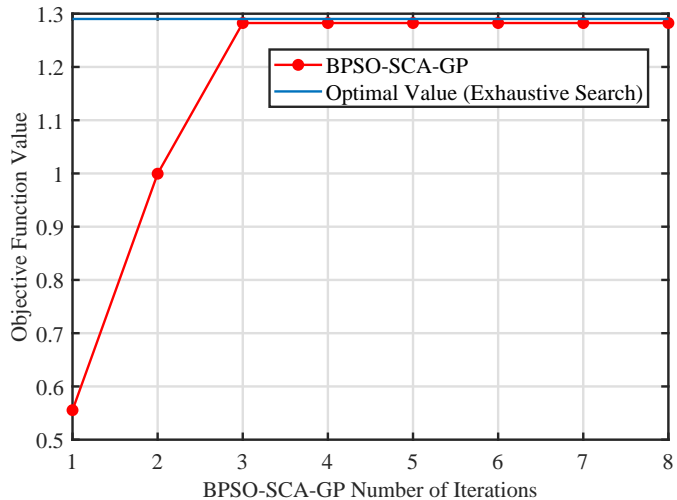}
		\caption{Exhaustive Search Vs. BPSO-SCA-GP}
		\label{fig:NearOptimalResult}
	\end{figure}

	Here, we have a performed an exhaustive search over all possible antenna selection combinations in a system where $M = 3,N = 2$ and $K = 3$. Of the $2$ antennas, one is considered to have a $1$ bit ADC and the other a high-resolution ADC. There can be $2^{MNK} = 2^{18} = 2,62,144$ feasible solutions for the antenna selection problem. For $E^{max} = 0.75E^{max}_{sys}$, many combinations does not fall into the feasible category and we are hence left with $1,14,624$ choices. However, this is still a huge number especially considering the network parameters. We calculate the power coefficient vector $\boldsymbol{\eta}$ for the possible choices of antenna selection coefficients which took roughly $300$ hours to complete, and determined the optimal value as shown in Fig. \ref{fig:NearOptimalResult}. Note that our algorithm reaches the optimal value in just four iterations in $140$ seconds.
	
	The simulation is performed on a desktop with Intel(R) Core(TM) i7-8700 CPU and running Windows 11. The clock of the machine is 3.20 GHz with a 32 GB memory.   The computational complexity of the BPSO-SCA-GP algorithm corresponds to the maximum number of times it needs to compute the objective function using the SCA-GP algorithm for power coefficient calculation. It depends on the maximum number of iteration, \textit{i.e.,} $I_{max}$ and the number of particle used in each iteration, \textit{i.e.,} $T$. Hence, the maximum number of times the BPSO-SCA-GP algorithm will call the SCA-GP subroutine is bounded by $I_{max} T$ whereas, in the case of exhaustive search, one needs to make $2^{MNK}$ such calls. In simulations, we considered $I_{max} = 50$ and $T = 10$ which is independent of other system parameters. Hence, the maximum number of times our algorithm call SCA-GP subroutine is bounded by $500$ which is much less than  $2^{MNK} = 2^{18} = 2,62,144$ which is the number of searches a exhaustive search has to make even for a toy system like $M = 3,N = 2$ and $K = 3$. Also, we have noticed that the BPSO algorithm typically converged in less than $10$ iteration, which means the actual calls for SCA-GP are below $100$.
	
	Next, we discussed the complexity of the SCA-GP algorithm. The complexity of the convex (GP) subproblem in the SCA-GP algorithm depends on the number of variables and constraints \cite{nasir2015joint}. In terms of big-$\mathcal{O}$ notation the complexity of GP subproblem is $\mathcal{O}\left( K^{4}\right)$ as there are $K$ variables and $K$ constraints. The overall complexity of SCA-GP can be obtained by multiplying the total number of iterations required for convergence ($N_{SCA}$) and the factor mentioned in terms of big-$\mathcal{O}$ notation. Finally, the complexity of BPSO-SCA-GP algorithm is bounded by $I_{max}T\left( N_{SCA} \mathcal{O}\left( K^{4}\right)\right)$ whereas the complexity of exhaustive search is $ 2^{MNK}\left( N_{SCA} \mathcal{O}\left( K^{4}\right)\right) $.
	
	\section{Conclusions}\label{sec:UCCF_Conclusion}
	This work studied a UC CF-mMIMO system, where APs are equipped with multiple antennas having a mixed ADC resolution profile. An algorithm for jointly optimizing the user's transmit power and antenna selection at the APs is proposed. An alternating optimization approach, utilizing the BPSO and SCA algorithms, is used to maximize the sum-rate of the system with constraints on the maximum energy consumed. Our simulation results demonstrate significant improvements in terms of SREE compared to schemes where a joint optimization is not performed. The effect of various system parameters such as number of APs, number of users, number of users served by an AP, the ratio of high-resolution to the total number of antennas is also studied. Some interesting future research directions include a) studying the performance improvements with multiple antennas at the users, b) coupling the presented architecture with a next-generation technology such as large intelligent reflecting surface (IRS) or non-orthogonal multiple access (NOMA) systems.

	\begin{appendices}
		\section{Proof for theorem \ref{Thm:UCCF_uplinkrate}}	
		\label{Proof:UCCF_uplinkrate}
		From (\ref{Eq:UCCF_skext}) and using the popular ``Use and forget" bound, the achievable uplink rate for $k$-th user can be expressed as follows 
		\begin{equation}
			\begin{aligned}\label{Eq:UCCF_UFExp}
				R_{k}^{UL} = \log_{2}\left(1 + \frac{\vert \mathrm{DS}_{k} \vert^{2}}{\mathbb{E}\left[\vert \mathrm{BU}_{k}\vert^{2}\right]+ \sum_{i \ne k}^{K} \mathbb{E}\left[\vert \mathrm{IUI}_{ki}\vert^{2} \right] + \mathbb{E}\left[\vert \mathrm{GN}_{k}\vert^{2}\right] + \mathbb{E}\left[\vert \mathrm{QN}_{k}\vert^{2}\right] } \right)
			\end{aligned}
		\end{equation}
		Next, we need to calculate the $\mathrm{DS}_{k}$, $\mathbb{E}\left[\vert \mathrm{BU}_{k}\vert^{2}\right]$, $\mathbb{E}\left[\vert \mathrm{IUI}_{ki}\vert^{2} \right]$, $\mathbb{E}\left[\vert \mathrm{GN}_{k}\vert^{2}\right]$ and $\mathbb{E}\left[\vert \mathrm{QN}_{k}\vert^{2}\right]$. First, we calculate $\mathrm{DS}_{k}$. Since, $\vec{g}_{mk} = \hat{\vec g}_{mk} + \tilde{\vec g}_{mk}$ where $\hat{\vec g}_{mk}$ is estimate and $\tilde{\vec g}_{mk}$ is error is estimation and both are independent so, we have     
		\begin{equation}
			\begin{aligned}\label{Eq:UCCF_DSk}
				\mathrm{DS}_{k}  &= \sqrt{\rho_{u}\eta_{k}}\mathds{E} \left[\sum_{m=1}^M \hat{\vec g}_{mk}^H \vec D_{mk} \vec A_m\left(\hat{\vec g}_{mk} + \tilde{\vec g}_{mk}\right)\right] \\
				&= \sqrt{\rho_{u}\eta_{k}}\sum_{m=1}^{M}\mathbb{E}\left[\operatorname{Tr}(\hat{\vec g}_{mk}^H \vec D_{mk} \vec A_m \hat{\vec g}_{mk})\right] \\
				&= \sqrt{\rho_{u}\eta_{k}}\sum_{m=1}^{M} \gamma_{mk} \operatorname{Tr}\left( \vec D_{mk} \vec A_m\right)
			\end{aligned}
		\end{equation}
		Second, we calculate $\mathbb{E}\left[\vert \mathrm{BU}_{k}\vert^{2}\right]$ and $\mathbb{E}\left[\vert \mathrm{IUI}_{k}\vert^{2}\right]$. From (\ref{Eq:UCCF_skext}), we have 
		\begin{equation}
			\begin{aligned}\label{Eq:UCCF_BUPower}
				\mathbb{E}\left[\vert \mathrm{BU}_{k}\vert^{2}\right] &=  \mathbb{E}\left[\abs*{ \sqrt{\rho_{u}\eta_{k}}\sum_{m=1}^{M} \hat{\vec g}_{mk}^H \vec D_{mk}\mathbf{A}_{m} \vec g_{mk}  - \mathrm{DS}_{k} }^{2}\right] \\
				&= \rho_{u}\eta_{k} \mathbb{E}\left[\abs*{ \sum_{m=1}^{M} \hat{\vec g}_{mk}^H \vec D_{mk}\mathbf{A}_{m} \vec g_{mk}}^{2}\right] - \abs*{\mathrm{DS}_{k}}^{2}
			\end{aligned}
		\end{equation}
		and 
		\begin{equation}
			\begin{aligned}\label{Eq:UCCF_IUIPower}
				\mathbb{E}\left[\vert \mathrm{IUI}_{k}\vert^{2}\right] &=  \mathbb{E}\left[\abs*{\sqrt{\rho_{u}}\sum_{i \ne k}^{K} \sum_{m=1}^M \sqrt{\eta_{i}}\hat{\vec g}_{mk}^H \vec D_{mk}\mathbf{A}_{m} \vec g_{mi}s_{i}}^{2}\right] \\
				&= \rho_{u}\sum_{i \ne k}^{K}\eta_{i} \mathbb{E}\left[\abs*{\sum_{m=1}^{M}\hat{\vec g}_{mk}^H \vec D_{mk}\mathbf{A}_{m} \vec g_{mi}}^{2}\right]
			\end{aligned}
		\end{equation}
		By adding (\ref{Eq:UCCF_BUPower}) and (\ref{Eq:UCCF_IUIPower}), we get
		\begin{equation}\label{Eq:UCCF_BUkplusIUIk}
			\begin{aligned}
				\mathbb{E}\left[\vert \mathrm{BU}_{k}\vert^{2}\right] + \mathbb{E}\left[\vert \mathrm{IUI}_{k}\vert^{2}\right] &= \rho_{u}\sum_{i = 1}^{K}\eta_{i} \mathbb{E}\left[\abs*{\sum_{m=1}^{M}\hat{\vec g}_{mk}^H \vec D_{mk}\mathbf{A}_{m} \vec g_{mi}}^{2}\right] - \abs*{\mathrm{DS}_{k}}^{2}
			\end{aligned}
		\end{equation}
		Now, we need $\mathbb{E}\left[ \abs*{\sum_{m=1}^M \hat{\vec g}_{mk}^H \vec D_{mk}  \vec A_m \vec g_{mi}}^{2}\right]$ which is equal to $\mathbb{E}\left[ \abs*{\sum_{m=1}^M \vec g_{mi}^H \vec A_m \vec D_{mk}   \hat{\vec g}_{mk}}^{2}\right]$. Using the fact that $\hat{\vec g}_{mk} = c_{mk}\left( \sqrt{\tau_{\mathrm{p}} \rho_{\mathrm{p}}} \sum_{j=1}^{K} \mathbf{g}_{m j}\boldsymbol{\phi}_{j}^{H}\boldsymbol{\phi}_{k} + \tilde{\mathbf{w}}_{\mathrm{p}, mk} \right)$, we have 
		\begin{equation}
			\begin{aligned}
				\hspace{-10mm}\mathbb{E}\left[ \abs*{\sum_{m=1}^M \vec g_{mi}^H \vec A_m \vec D_{mk}   \hat{\vec g}_{mk}}^{2}\right] = \mathbb{E}\left[ \abs*{\sum_{m=1}^M \vec g_{mi}^H \vec A_m \vec D_{mk}   c_{mk}\left( \sqrt{\tau_{\mathrm{p}} \rho_{\mathrm{p}}} \sum_{j=1}^{K} \mathbf{g}_{m j}\boldsymbol{\phi}_{j}^{H}\boldsymbol{\phi}_{k} + \tilde{\mathbf{w}}_{\mathrm{p}, mk} \right)}^{2}\right]
			\end{aligned}
		\end{equation}
		where, $\tilde{\mathbf{w}}_{\mathrm{p}, mk} \sim \mathcal{CN}\left(\mathbf{0}_{N},\mathbf{I}_{N}\right) $, is independent of $ \vec{g}_{mi} \ \forall i $, so we have 
		\begin{equation}\label{Eq:UCCF_IUImidterm}
			\begin{aligned}
				\mathbb{E}\left[ \Bigg\vert\sum_{m=1}^M \vec g_{mi}^H \vec A_m \vec D_{mk}   \hat{\vec g}_{mk}\Bigg\vert^{2}\right]  &= \tau_{p} \rho_{p}\underbrace{\mathbb{E}\left[ \Bigg\vert \sum_{m=1}^M \sum_{j=1}^{K} \left(c_{mk} \boldsymbol{\phi}_{j}^{H}\boldsymbol{\phi}_{k}\right)\vec g_{mi}^H \vec D_{mk}  \vec A_m   \mathbf{g}_{m j}\Bigg\vert^{2}\right]}_{T_{1}}
				\\& + \underbrace{\mathbb{E}\left[ \Bigg\vert\sum_{m=1}^M \vec g_{mi}^H \vec D_{mk}  \vec A_m c_{mk}\tilde{\mathbf{w}}_{\mathrm{p}, mk} \Bigg\vert^{2}\right]}_{T_{2}}
			\end{aligned}
		\end{equation}
		First, we compute $T_{1}$. We have 
		\begin{equation}
			\begin{aligned}
				T_{1} &= \mathbb{E}\left[ \Bigg\vert \sum_{m=1}^M \sum_{j=1}^{K} \left(c_{mk} \boldsymbol{\phi}_{j}^{H}\boldsymbol{\phi}_{k}\right)\vec g_{mi}^H \vec D_{mk}  \vec A_m   \mathbf{g}_{m j}\Bigg\vert^{2}\right] \\
				&= \mathbb{E}\left[  \sum_{m=1}^M \sum_{j=1}^{K} \sum_{n=1}^M \sum_{t=1}^{K} \left(c_{mk} \boldsymbol{\phi}_{j}^{H}\boldsymbol{\phi}_{k}\right) \left(c_{nk} \boldsymbol{\phi}_{k}^{H}\boldsymbol{\phi}_{t}\right) \operatorname{Tr}\left(  \vec D_{mk}  \vec A_m   \mathbf{g}_{m j} \mathbf{g}_{n t}^{H} \vec A_{n} \vec D_{nk} \mathbf{g}_{ni} \vec g_{mi}^H\right) \right]
			\end{aligned}
		\end{equation}
		After taking the expectation inside the summation and some algebraic manipulations, we have
		\begin{equation}\label{Eq:UCCF_T1cal}
			\begin{aligned}
				T_{1} &= \sum_{m=1}^M  \sum_{n=1}^M  c_{mk} c_{nk} \vert \boldsymbol{\phi}_{k}^{H}\boldsymbol{\phi}_{i} \vert^{2} \beta_{mi}\beta_{ni} \operatorname{Tr}\left(\vec D_{mk}\vec A_{m}\right) \operatorname{Tr}\left(\vec D_{nk}\vec A_{n}\right) \\&+ \sum_{m=1}^M \sum_{j=1}^{K} c_{mk}^{2} \vert\boldsymbol{\phi}_{k}^{H}\boldsymbol{\phi}_{j} \vert^{2} \beta_{mi} \beta_{mj}\operatorname{Tr}\left(\vec D_{mk} \vec A_{m}^{2}\right)
			\end{aligned}
		\end{equation}
		Similarly, we have
		\begin{equation}\label{Eq:UCCF_T2cal}
			\begin{aligned}
				T_{2} &= \mathbb{E}\left[ \sum_{m=1}^M c_{mk}^{2} \operatorname{Tr} \left( \vec g_{mi}^H \vec D_{mk}  \vec A_m \tilde{\mathbf{w}}_{\mathrm{p}, mk}\tilde{\mathbf{w}}_{\mathrm{p}, mk}^{H}\vec A_m \vec D_{mk} \vec g_{mi} \right) \right] \\
				&= \mathbb{E}\left[ \sum_{m=1}^M c_{mk}^{2} \operatorname{Tr} \left(  \vec D_{mk}  \vec A_m \tilde{\mathbf{w}}_{\mathrm{p}, mk}\tilde{\mathbf{w}}_{\mathrm{p}, mk}^{H}\vec A_m \vec D_{mk} \vec g_{mi} \vec g_{mi}^H \right) \right] \\
				&=  \sum_{m=1}^M c_{mk}^{2}\beta_{mi} \operatorname{Tr} \left(  \vec D_{mk}  \vec A_{m}^{2}  \right)
			\end{aligned}
		\end{equation}
		Finally, by substituting (\ref{Eq:UCCF_T1cal}) and (\ref{Eq:UCCF_T2cal}) in (\ref{Eq:UCCF_IUImidterm}), we have
		\begin{equation}\label{Eq:UCCF_IUImidtermFinal}
			\begin{aligned}
				\mathbb{E}\left[ \abs*{\sum_{m=1}^M \vec g_{mi}^H \vec D_{mk}  \vec A_m \hat{\vec g}_{mk}}^{2}\right] &= \vert \boldsymbol{\phi}_{k}^{H}\boldsymbol{\phi}_{i} \vert^{2} \left( \sum_{m=1}^M    \gamma_{mk}   \frac{\beta_{mi}}{\beta_{mk}} \operatorname{Tr}\left(\vec D_{mk}\vec A_{m}\right)\right)^{2} \\&+  \sum_{m=1}^M \gamma_{mk}\beta_{mi} \operatorname{Tr} \left(  \vec D_{mk}  \vec A_{m}^{2}  \right)
			\end{aligned}
		\end{equation}
		and substitution of (\ref{Eq:UCCF_IUImidtermFinal}) and (\ref{Eq:UCCF_DSk}) results in 
		\begin{equation}\label{Eq:UCCF_BU_IUI}
			\begin{aligned}
				\mathbb{E}\left[\vert \mathrm{BU}_{k}\vert^{2}\right] + \mathbb{E}\left[\vert \mathrm{IUI}_{k}\vert^{2}\right] &= \rho_{u}\sum_{i \ne k}^{K}{\eta_i} \left( \sum_{m=1}^M    \gamma_{mk}   \frac{\beta_{mi}}{\beta_{mk}} \operatorname{Tr}\left(\vec D_{mk}\vec A_{m}\right)\right)^{2} \vert \boldsymbol{\phi}_{k}^{H}\boldsymbol{\phi}_{i} \vert^{2} \\ &+ \sum_{m=1}^M \gamma_{mk}\beta_{mi} \operatorname{Tr} \left(  \vec D_{mk}  \vec A_{m}^{2}  \right)
			\end{aligned}
		\end{equation}
		Next, we compute $\mathbb{E}\left[\vert \mathrm{GN}_{k}\vert^{2}\right]$. From (\ref{Eq:UCCF_skext}), we have $\mathrm{GN}_{k} = \sum_{m=1}^M \hat{\vec g}_{mk}^H \vec D_{mk}\mathbf{A}_{m}\vec w_{m}$, hence
		\begin{equation}
			\begin{aligned}
				\mathbb{E}\left[\vert \mathrm{GN}_{k}\vert^{2}\right] = \mathbb{E}\left[\Big\vert \sum_{m=1}^M \hat{\vec g}_{mk}^H \vec D_{mk}\mathbf{A}_{m}\vec w_{m}\Big\vert^{2}\right]
			\end{aligned}
		\end{equation}
		Using the fact that $\hat{\vec g}_{mk}^H$ and $\vec{w}_{m}$ are independent, we have 
		\begin{equation}\label{Eq:UCCF_GNk}
			\begin{aligned}
				\mathbb{E}\left[\vert \mathrm{GN}_{k}\vert^{2}\right] &= \mathbb{E}\left[\sum_{m=1}^{M}\norm*{\vec A_m  \vec D_{mk} \hat{\vec g}_{mk}}^{2}\right] \\
				&= \sum_{m=1}^{M} \mathbb{E}\left[\hat{\vec g}_{mk}^H \vec D_{mk} \vec A_m \vec A_m \vec D_{mk} \hat{\vec g}_{mk} \right] \\
				&= \sum_{m=1}^{M} \gamma_{mk} \operatorname{Tr}\left( \vec D_{mk} \vec A_m^2\right)
			\end{aligned}
		\end{equation}
		Finally, we compute $\mathbb{E}\left[\vert \mathrm{QN}_{k}\vert^{2}\right]$. From (\ref{Eq:UCCF_skext}), we have $\mathrm{QN}_{k} = \sum_{m=1}^{M} \hat{\vec g}_{mk}^H \vec D_{mk} \vec w_{m}^q $, hence
		\begin{equation}
			\begin{aligned}
				\mathbb{E}\left[\vert \mathrm{QN}_{k}\vert^{2}\right] &=  \mathbb{E}\left[\abs*{\sum_{m=1}^{M} \hat{\vec g}_{mk}^H \vec D_{mk} \vec w_{m}^q}^2 \right] \\
				&= \sum_{m=1}^{M} \mathbb{E}\left[ \vert \hat{\vec g}_{mk}^H \vec D_{mk} \vec w_{m}^q \vert^2\right] 
			\end{aligned}
		\end{equation}
		For a given channel realization, we have $$ \vec w_{m}^{q} \sim \mathcal{CN}\left(\vec 0,  \mathbf{A}_{m} \left(\mathbf{I}_{N} - \mathbf{A}_{m}\right)\operatorname{diag}\left(\rho_{u}\vec G_{m}\vec P \vec G_{m}^{H} + \mathbf{I}_{N}\right)\right) $$
		\begin{equation}
			\begin{split}
				\mathbb{E}\left[ \vert \hat{\vec g}_{mk}^H \vec D_{mk} \vec w_{m}^q \vert^2\right] &= \mathbb{E}\left[  \hat{\vec g}_{mk}^H \vec D_{mk} \vec w_{m}^q \vec {w_{m}^q}^{H} \vec D_{mk} \hat{\vec g}_{mk} \right]\\
				&= \mathbb{E}\left[ \hat{\vec g}_{mk}^H \vec D_{mk} \mathbf{A}_{m} \left(\mathbf{I}_{N} - \mathbf{A}_{m}\right)\operatorname{diag}\left(\rho_{u}\vec G_{m}\vec P \vec G_{m}^{H} + \mathbf{I}_{N}\right) \vec D_{mk} \hat{\vec g}_{mk} \right] \\
				&= \mathbb{E}\left[\operatorname{Tr}\left(  \vec D_{mk} \mathbf{A}_{m} \left(\mathbf{I}_{N} - \mathbf{A}_{m}\right)\operatorname{diag}\left(\rho_{u}\vec G_{m}\vec P \vec G_{m}^{H} + \mathbf{I}_{N}\right) \hat{\vec g}_{mk}\hat{\vec g}_{mk}^H \right) \right]
			\end{split}
		\end{equation}
		We got the above equation using the cyclic property of trace function and commutative nature of multiplication of diagonal matrices. Further, we can simplify as follows
		\begin{equation}\label{Eq:UCCF_quatizationnoise}
			\begin{aligned}
				\mathbb{E}\left[ \vert \hat{\vec g}_{mk}^H \vec D_{mk} \vec w_{m}^q \vert^2\right] &= \mathbb{E}\left[\operatorname{Tr}\left(  \vec D_{mk} \mathbf{A}_{m} \left(\mathbf{I}_{N} - \mathbf{A}_{m}\right)\lbrace \operatorname{diag}\left(\rho_{u}\vec G_{m}\vec P \vec G_{m}^{H}\right) \hat{\vec g}_{mk}\hat{\vec g}_{mk}^H + \hat{\vec g}_{mk}\hat{\vec g}_{mk}^H \rbrace \right) \right] \\
				&= \operatorname{Tr}\left(  \vec D_{mk} \mathbf{A}_{m} \left(\mathbf{I}_{N} - \mathbf{A}_{m}\right)\mathbb{E}\left[ \operatorname{diag}\left(\rho_{u} \vec G_{m}\vec P \vec G_{m}^{H}\right) \hat{\vec g}_{mk}\hat{\vec g}_{mk}^H + \hat{\vec g}_{mk}\hat{\vec g}_{mk}^H \right] \right) \\
				&= \rho_{u}\operatorname{Tr}\left(  \vec D_{mk} \mathbf{A}_{m} \left(\mathbf{I}_{N} - \mathbf{A}_{m}\right)\mathbb{E}\left[ \operatorname{diag}\left(\vec G_{m}\vec P \vec G_{m}^{H}\right) \hat{\vec g}_{mk}\hat{\vec g}_{mk}^H\right]\right) \\&\quad+ \gamma_{mk}\operatorname{Tr}\left(\vec D_{mk} \mathbf{A}_{m} \left(\mathbf{I} - \mathbf{A}_{m}\right)\right) 
			\end{aligned}
		\end{equation}
		Now we need to solve $\mathbb{E}\left[ \operatorname{diag}\left(\vec G_{m}\vec P \vec G_{m}^{H}\right) \hat{\vec g}_{mk}\hat{\vec g}_{mk}^H\right]$
		\begin{equation}
			\begin{aligned}
				\mathbb{E}\left[ \operatorname{diag}\left(\vec G_{m}\vec P \vec G_{m}^{H}\right) \hat{\vec g}_{mk}\hat{\vec g}_{mk}^H\right] &= \mathbb{E}\left[ \operatorname{diag}\left(\sum_{i=1}^{K}\eta_{i} \vec g_{mi}\vec g_{mi}^{H} \right) \hat{\vec g}_{mk}\hat{\vec g}_{mk}^H\right] \\
				&= \mathbb{E}\left[ \sum_{i=1}^{K}\eta_{i} \operatorname{diag}\left(\vec g_{mi}\vec g_{mi}^{H} \right) \hat{\vec g}_{mk}\hat{\vec g}_{mk}^H\right]
			\end{aligned}
		\end{equation}
		
		\begin{equation}
			\begin{aligned}
				\because     \operatorname{diag}\left(\vec g_{mi} \vec g_{mi}^{H} \right) = \begin{bmatrix}
					\vert \vec g_{mi}^{1} \vert^{2} &  & \\
					& \ddots &  \\
					&  & \vert \vec g_{mi}^{N} \vert^{2} 
				\end{bmatrix}  \text{and} \ \hat{\vec g}_{mk}\hat{\vec g}_{mk}^H = \begin{bmatrix}
					\vert \hat{\vec g}_{mk}^{1} \vert^{2} & \hat{\vec g}_{mk}^{1}\left(\hat{\vec g}_{mk}^{2}\right)^{*}   & \cdots & \hat{\vec g}_{mk}^{1}\left(\hat{\vec g}_{mk}^{N}\right)^{*}\\
					\vdots &  &  & \vdots\\
					\hat{\vec g}_{mk}^{N}\left(\hat{\vec g}_{mk}^{1}\right)^{*} & \hat{\vec g}_{mk}^{N}\left(\hat{\vec g}_{mk}^{2}\right)^{*} &  \cdots &\vert \hat{\vec g}_{mk}^{N} \vert^{2} 
				\end{bmatrix}
			\end{aligned}\nonumber
		\end{equation}
		Hence, we have 
		\begin{equation}
			\begin{aligned}
				&\mathbb{E}\left[ \sum_{i=1}^{K}\eta_{i} \operatorname{diag}\left(\vec g_{mi}\vec g_{mi}^{H} \right) \hat{\vec g}_{mk}\hat{\vec g}_{mk}^H\right] = \\ &\qquad \qquad \quad\mathbb{E}\left[\sum_{i=1}^{K}\eta_{i} \begin{bmatrix} 
					\vert \vec g_{mi}^{1} \vert^{2} \vert \hat{\vec g}_{mk}^{1} \vert^{2} & \vert \vec g_{mi}^{1} \vert^{2} \hat{\vec g}_{mk}^{1}\left(\hat{\vec g}_{mk}^{2}\right)^{*}   & \cdots & \vert \vec g_{mi}^{1} \vert^{2} \hat{\vec g}_{mk}^{1}\left(\hat{\vec g}_{mk}^{N}\right)^{*}\\
					\vdots &  &  & \vdots\\
					\vert \vec g_{mi}^{N} \vert^{2} \hat{\vec g}_{mk}^{N}\left(\hat{\vec g}_{mk}^{1}\right)^{*} & \vert \vec g_{mi}^{N} \vert^{2}\hat{\vec g}_{mk}^{N}\left(\hat{\vec g}_{mk}^{2}\right)^{*} &  \cdots & \vert \vec g_{mi}^{N} \vert^{2} \vert \hat{\vec g}_{mk}^{N} \vert^{2} 
				\end{bmatrix} \right]
			\end{aligned}
		\end{equation}
		$\because \mathbb{E}\left[ \hat{\vec g}_{mk} \hat{\vec g}_{mk}^{H} \right]$ is diagonal $\therefore$ estimate at different antenna will be uncorrelated. So the expectation of all the off-diagonal terms will be zero. Hence we have
		\begin{equation}
			\begin{aligned}
				\mathbb{E}\left[ \sum_{i=1}^{K}\eta_{i} \operatorname{diag}\left(\vec g_{mi}\vec g_{mi}^{H} \right) \hat{\vec g}_{mk}\hat{\vec g}_{mk}^H\right] = \sum_{i=1}^{K}\eta_{i} \begin{bmatrix} 
					\mathbb{E}\left[ \vert \vec g_{mi}^{1} \vert^{2} \vert \hat{\vec g}_{mk}^{1} \vert^{2}\right] &  & \\
					& \ddots &  \\
					&  & \mathbb{E}\left[ \vert \vec g_{mi}^{N} \vert^{2} \vert \hat{\vec g}_{mk}^{N} \vert^{2} \right] 
				\end{bmatrix}
			\end{aligned}
		\end{equation}
		Now we need to calculate the $\mathbb{E}\left[ \vert \vec g_{mi}^{1} \vert^{2} \vert \hat{\vec g}_{mk}^{1} \vert^{2}\right] $. As $\hat{\vec g}_{mk} = c_{mk}\left( \sqrt{\tau_{\mathrm{p}} \rho_{\mathrm{p}}} \sum_{j=1}^{K} \mathbf{g}_{m j}\boldsymbol{\phi}_{j}^{H}\boldsymbol{\phi}_{k} + \tilde{\mathbf{w}}_{\mathrm{p}, mk} \right)$ and $\vec{\hat{g}}_{mk} \sim \mathcal{CN}\left(\vec 0, \gamma_{mk}\mathbf{I}_{N}\right)$. Note that the $\vec{g}_{mi}$ and $\hat{\vec g}_{mk}$ both are vectors of i.i.d. random variables so $\mathbb{E}\left[ \vert \vec g_{mi}^{n} \vert^{2} \vert \hat{\vec g}_{mk}^{n} \vert^{2}\right] $ will be same for all $n = 1,\dots,N$
		
		\begin{equation}
			\begin{aligned}
				\mathbb{E}\left[ \vert \vec g_{mi}^{n} \vert^{2} \vert \hat{\vec g}_{mk}^{n} \vert^{2}\right] &= \gamma_{mk}\beta_{mi} + \gamma_{mk}\gamma_{mi} \vert \boldsymbol{\phi}_{k}^{H}\boldsymbol{\phi}_{i} \vert^{2} 
			\end{aligned}
		\end{equation}
		so,
		\begin{equation}
			\begin{aligned}
				\mathbb{E}\left[ \sum_{i=1}^{K}\eta_{i} \operatorname{diag}\left(\vec g_{mi}\vec g_{mi}^{H} \right) \hat{\vec g}_{mk}\hat{\vec g}_{mk}^H\right] = \gamma_{mk}\left(\sum_{i=1}^{K}\eta_{i} \left(\beta_{mi} + \gamma_{mi} \vert \boldsymbol{\phi}_{k}^{H}\boldsymbol{\phi}_{i} \vert^{2} \right) \right) \mathbf{I}_{N}.
			\end{aligned}
		\end{equation}	
		After putting the above equation in (\ref{Eq:UCCF_quatizationnoise}), we have 
		\begin{equation}\label{Eq:UCCF_QNk}
			\begin{aligned}
				\mathbb{E}\left[ \vert \hat{\vec g}_{mk}^H \vec D_{mk} \vec w_{m}^q \vert^2\right] 
				&= \rho_{u}\operatorname{Tr}\left(  \vec D_{mk} \mathbf{A}_{m} \left(\mathbf{I}_{N} - \mathbf{A}_{m}\right)\mathbb{E}\left[ \operatorname{diag}\left(\vec G_{m}\vec P \vec G_{m}^{H}\right) \hat{\vec g}_{mk}\hat{\vec g}_{mk}^H\right]\right) \\&\quad+ \gamma_{mk}\operatorname{Tr}\left(\vec D_{mk} \mathbf{A}_{m} \left(\mathbf{I} - \mathbf{A}_{m}\right)\right) \\
				&= \rho_{u}\gamma_{mk}\left(\sum_{i=1}^{K}\eta_{i} \left(\beta_{mi} + \gamma_{mi} \vert \boldsymbol{\phi}_{k}^{H}\boldsymbol{\phi}_{i} \vert^{2} \right) \right)\operatorname{Tr}\left(\vec D_{mk} \mathbf{A}_{m} \left(\mathbf{I} - \mathbf{A}_{m}\right)\right) \\&\quad+ \gamma_{mk}\operatorname{Tr}\left(\vec D_{mk} \mathbf{A}_{m}\right) - \gamma_{mk}\operatorname{Tr}\left(\vec D_{mk} \mathbf{A}_{m}^{2} \right) 
			\end{aligned}
		\end{equation}
		The result in (\ref{Eq:UCCF_UPRatek}) follows by substituting (\ref{Eq:UCCF_DSk}),(\ref{Eq:UCCF_BU_IUI}),(\ref{Eq:UCCF_GNk}) and (\ref{Eq:UCCF_QNk}) in (\ref{Eq:UCCF_UFExp}), and this completes the proof.
	\end{appendices}
	
	\bibliographystyle{IEEEtran}
	\bibliography{UC_Cell_Free_ADC_Version_2}
\end{document}